\documentclass[twocolumn,superscriptaddress,preprintnumbers,amsmath,amssymb,floatfix,nofootinbib]{revtex4-1}
\usepackage{graphicx}
\usepackage{amsmath}
\usepackage{amssymb}
\usepackage{color}
\usepackage{cases}
\usepackage{soul}

\begin{document}

\title{Bayesian approach to SETI}

\author{Claudio Grimaldi}\email{claudio.grimaldi@epfl.ch}
\affiliation{Laboratory of Physics of Complex Matter, Ecole Polytechnique F\'ed\'erale
de Lausanne, Station 3, CP-1015 Lausanne, Switzerland}
\author{Geoffrey W. Marcy}\email{geoff.w.marcy@gmail.com}
\affiliation{University of California, Berkeley, CA 94720, USA}


\begin{abstract}
The search for technosignatures from hypothetical galactic civilizations is going through a new phase of intense activity.
For the first time, a significant fraction of the vast search space is expected to be sampled in the foreseeable future, 
potentially bringing informative data about the abundance of detectable extraterrestrial civilizations, or the lack thereof.
Starting from the current state of ignorance about the galactic population of non-natural electromagnetic signals, we formulate a
Bayesian statistical model to infer the mean number of radio signals crossing Earth, assuming either non-detection
or the detection of signals in future surveys of the Galaxy. 
Under fairly noninformative priors, we find that not detecting signals within about $1$ kly from Earth, while suggesting
the lack of galactic emitters or at best the scarcity thereof, is nonetheless still 
consistent with a probability exceeding $10$ \% that typically over $\sim 100$ signals could be crossing Earth, with
radiated power analogous to that of the Arecibo radar, but coming from farther in the Milky Way.
The existence in the Galaxy of potentially detectable Arecibo-like emitters can be reasonably ruled out only if all-sky surveys detect no such signals 
up to a radius of about $40$ kly, an endeavor requiring detector sensitivities thousands times higher than those of current telescopes. Conversely,
finding even one Arecibo-like signal within $\sim 1000$ light years, a possibility within reach of current detectors, implies almost 
certainly that typically more than $\sim 100$ signals of comparable radiated power cross the Earth, yet to be discovered.
\end{abstract}

\maketitle

\section{Introduction}
\label{intro}
SETI, the search for extraterrestrial intelligence pursued primarily by seeking non-natural electromagnetic (EM) signals in the 
Galaxy, is notoriously a challenging endeavor with unknown chances of success. Because of the small fraction of the SETI search space explored 
so far, the non-detections to date of non-natural signals contain only modest informative value about the existence of extraterrestrial
technological civilizations in the entire Milky Way. For example, the most recent targeted search for radio transmissions detected no signals in the 
frequency range between $1.1$ and $1.9$ GHz from $692$ nearby stars, suggesting that fewer than $\sim 0.1$ \% of stars within $\sim 160$ ly 
harbor transmitters whose signals cross the Earth and having equivalent isotropic radiated power
comparable to or larger than that of terrestrial planetary radars \cite{Enriquez2017}. This fraction drops to about $0.01$ \% if 
emitters are assumed to transmit uniformly between $\sim 1$ and $\sim 10$ GHz, the frequency range defining the terrestrial microwave window
thought to give the best opportunity to detect non-natural EM signals. Extrapolating this result to the entire Galaxy gives a vivid picture
of our current state of ignorance. An upper limit of $0.1$\%-$0.01$\% of stars possessing detectable emitters is indeed consistent
with the Earth being illuminated by a total number of radio signals ranging from $0$ to $10^6$-$10^7$, even if we consider only sun-like stars with
Earth-size planets \cite{Petigura2013}.

This state of extreme uncertainty may however change. The discovery of thousands of extrasolar planets \cite{Batalha2013} and the 
inferred astronomical number of Earth-like planets in the Galaxy\cite{Petigura2013} have recently stimulated a significant revival of SETI initiatives. 
The ``Breakthrough Listen" project \cite{Isaacson2017,Enriquez2017}, the largest and most
comprehensive search ever, and the planned ``Cradle of Life" program of the Square Kilometre Array radiotelescope \cite{Lazio2004,Siemion2015}, 
together with impressive progress in detector technology \cite{Loeb2007,Garrett2017}, offer unprecedented opportunities for a systematic 
investigation in the vast domain of the SETI search space.

In view of these rapid developments, exploration of a significant fraction of the Galaxy for a broad range of wavelengths has to be expected in the
following years, providing a sufficiently large amount of informative data to infer, at least to some extent, the possible galactic population
of non-natural, extraterrestrial signals in the Milky Way.

Here, we report the results of a bayesian analysis formulated by assuming either non-detection or the detection of a signal 
within a given radio frequency range as a function of the volume of the Galaxy sampled by an hypothetical SETI survey. 
We construct a statistical model that considers possible populations of  
extraterrestrial emitters, their spatial and age distributions, and the longevity of the emission processes. By taking into account the luminosity
distribution of the emitters and the sensitivities of the detectors, we calculate the posterior probabilities of the 
average number of signals crossing Earth emitted from the entire Milky Way, given the present very limited level of knowledge.
The results show that not detecting signals out to a distance of about $40$ kly from Earth places a strong upper limit on the occurrence of
detectable EM emissions from the entire Galaxy. This limit can be reached by with future radio telescopes, such as the Phase 2 of the 
Square Kilometre Array if emitters more powerful than terrestrial planetary radars are assumed.
In contrast, the detection of even a single signal from the galactic neighborhood (i.e., within a distance of $\sim 1$ kly from Earth) hints to
a posterior probability of almost $100$ \% that hundreds of signals from the entire Galaxy typically cross the Earth, with an even larger total 
number of signals populating the Galaxy.

While our analysis focuses here on radio signals, the formalism can be extended to consider other wavelengths, like the optical and near
infrared spectrum searched by some SETI initiatives \cite{Townes1983,Tellis2015,Tellis2017,Wright2018}. 
In the case of short wavelengths, however, absorption and scattering processes have 
to be considered, as briefly discussed in the concluding section.

\section{The Model}
\label{model}
In modeling possible galactic populations of non-natural extraterrestrial signals, we start by considering an hypothetical technological,
communicating civilization (or emitter) located at some position vector $\vec{r}$ relative to the galactic center.
We assume that at some time $t$ in the past the emitter started transmitting, either deliberately or not, an isotropic EM
signal, and that the emission process lasted a time interval denoted $L$. 
At the present time
the region of space occupied by the EM radiation is a spherical shell centered at $\vec{r}$, with outer radius $R=ct$ and thickness 
$\Delta=cL$, where $c$ is the speed of light.

\begin{figure}[t]
	\begin{center}
		\includegraphics[scale=0.2,clip=true]{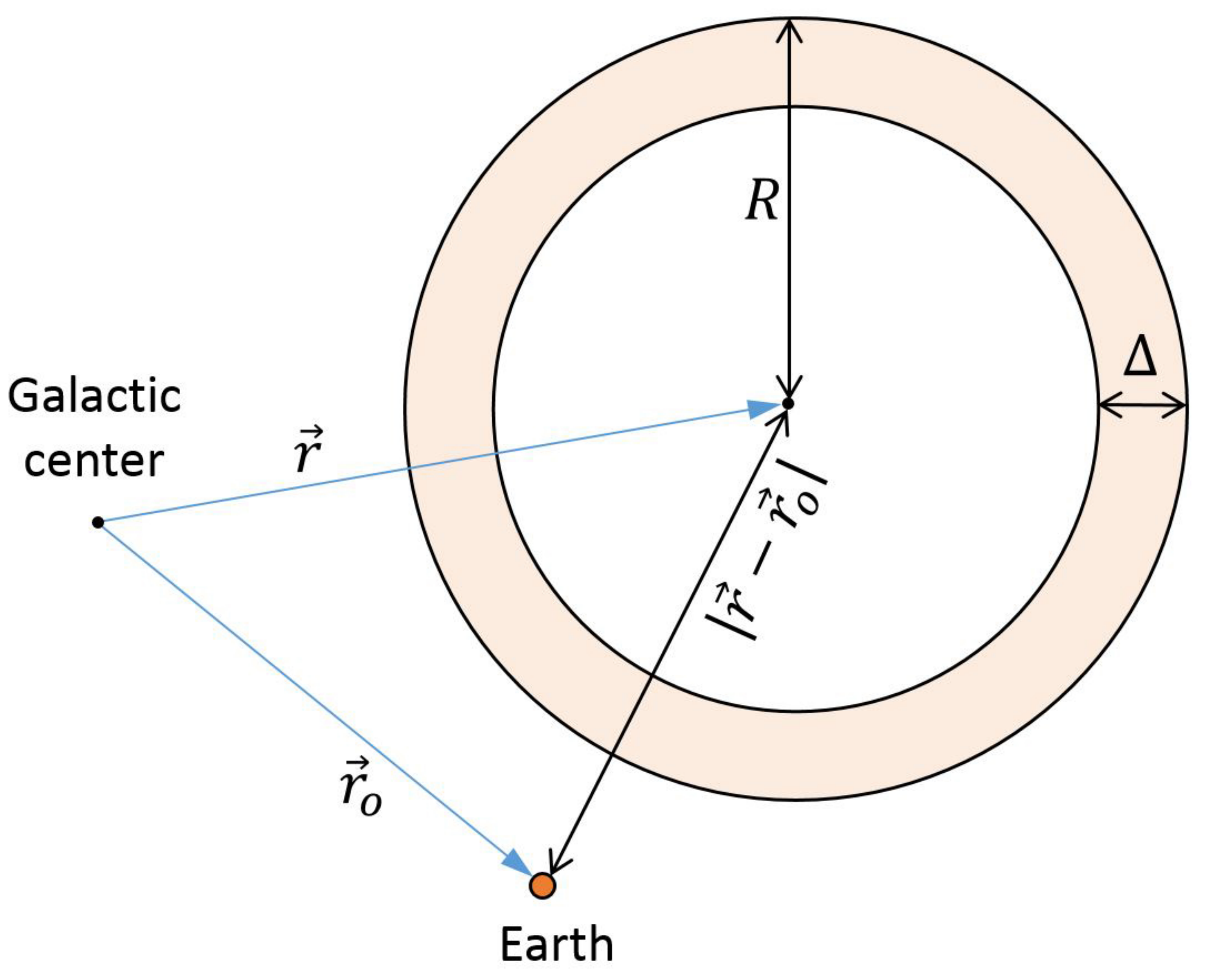}
		\caption{Two-dimensional schematic representation of a spherical shell signal of outer radius $R$ and thickness $\Delta$.
			The spherical shell is centered at the emitter location identified by the position vector $\vec{r}$ relative to the
			galactic center, while the Earth (red circle) is located at $\vec{r}_o$. In the figure, the Earth
			lies outside the region covered by the shell, preventing the detection of the signal. The spherical shell signal intercepts
			the Earth only if the distance emitter-Earth,
			$\vert\vec{r}-\vec{r}_o\vert$, satisfies Eq.~\ref{cond1}.
		}\label{fig1}
	\end{center}
\end{figure}

A necessary condition for the detection of this signal is that, at the time of observation, the position vector of the Earth, $\vec{r}_o$,
points to a location within the region occupied by the spherical shell, which corresponds to requiring that \cite{Grimaldi2017,Grimaldi2018}
\begin{equation}
\label{cond1} 
R-\Delta\leq \vert\vec{r}-\vec{r}_o\vert\leq R,
\end{equation}
where $\vert\vec{r}-\vec{r}_o\vert$ is the distance of the emitter from the Earth, Fig.~\ref{fig1}. The first inequality of Eq.~\ref{cond1} represents 
the condition that the last emitted signal of a spherical shell crosses Earth \cite{Balbi2018}.
Since the farthest possible position of a galactic emitter is at the opposite edge of the galactic disk, its maximum conceivable distance
from the Earth, $R_M\approx 87$ kly, is simply the sum of the galactic radius ($\approx 60$ kly\footnote[1]{
	Here we adopt the presumption that stars that can potentially harbor emitters are rich in heavy elements, so to favor the formation of rocky 
	planets. $60$ kly is approximately the radius of the galactic thin disk (see Sec.~\ref{bayanalysis}) which is a metal-rich component of the Milky Way.
	The contribution of farther stars \cite{deGeus1993,Anderson2015} and/or other galactic components \cite{DiStefano2016} can nonetheless be 
	incorporated by the present formalism})
and the distance of the Earth from the galactic center ($\vert\vec{r}_o\vert\approx 27$ kly). Hence, any EM signal emitted before 
$t_M=R_M/c\approx 87,000$ years from present has already covered a distance larger than $R_M$, and is therefore absolutely undetectable at Earth. 
Since $\vert\vec{r}-\vec{r}_o\vert<R_M$, it follows also that the region filled by a spherical shell with outer radius larger 
than $R_M+\Delta$ cannot contain our planet, regardless of the position of the emitter in the Milky Way. 
In temporal terms, this means that any emission process lasting $L$ years and that started at a time earlier than $t_M+L$ is
unobservable and can be ignored.

\subsection{Probability of shells at Earth}
\label{pse}

To calculate the probability of signals crossing Earth, we must consider all possible configurations of the spherical shells
(their number, position, outer radius and thickness) and identify those that satisfy Eq.~\ref{cond1}. To this end, we first
discard any unobservable signal by assigning to a randomly chosen star 
a probability $q$ of harboring an emitter that has been actively transmitting some time within $t_M$
years from present. The mean number of such emitters is thus $qN_s$, where $N_s$ is the number of stars in the Galaxy.
We make the additional assumption that the starting time and the duration of the emissions (or, equivalently, the outer radius and 
the thickness of the spherical shells) are independent and identically distributed random variables, $t$ and $L$, with 
probability density functions (PDFs) given by $\rho_t(t)$ and $\rho_L(L)$, respectively, and that the signal
frequencies are distributed uniformly within a given range. The resulting probability $p$ that the Earth intersects a signal 
under the condition that it is no older than $t_M$ (or, equivalently, that the emission process started 
within a time $t_M+L$ before present) is therefore \cite{Grimaldi2017}
\begin{equation}
\label{p1}
p=q\frac{\displaystyle\int\!\!d L\,\rho_L(L)\!\int_0^{t_M+L}\!\!dt\,\rho_t(t)\!\int\!\!d\vec{r}\rho_s(\vec{r})
	f_{R,\Delta}(\vec{r}-\vec{r}_o)}
{N_s\displaystyle\int\!\!d L\,\rho_L(L)\!\int_0^{t_\textrm{M}+L}\!\!dt\,\rho_t(t)},
\end{equation}
where $f_{R,\Delta}(\vec{r}-\vec{r}_o)=1$ if Eq.~\ref{cond1} is satisfied and $f_{R,\Delta}(\vec{r}-\vec{r}_o)=0$ 
otherwise, and $\rho_s(\vec{r})$ is the star number density function. For the moment, we do not need to specify its 
detailed form and require only that $\int\!d\vec{r}\rho_s(\vec{r})=N_s$ and that 
$\rho_s(\vec{r})$ has approximately a disk-like shape with a radius of about $60$ kly.

Assuming the steady-state condition that over a time span of order $t_M$ from present the PDF of the starting 
time of emission, $\rho_t(t)$, is essentially constant \cite{Grimaldi2017}, the integrals over $t$ in Eq.~\ref{p1} 
can be solved exactly and $p$ reduces simply to (SI Appendix, Section I): 
\begin{equation}
\label{lambda}
p=q\lambda\equiv q\frac{\bar{L}}{\bar{L}+t_M},
\end{equation}
where the second equality defines the scaled longevity of the signal, $\lambda$, and $\bar{L}=\int\! dL\rho_L(L)L$ denotes
the average duration of the signal. Finally, from Eq.~\ref{lambda} we obtain the mean number of signals
crossing Earth,
\begin{equation}
\label{kbar}
\bar{k}=q\lambda N_s,
\end{equation}
which has to be understood as a statistical average over all configurations of the emitted signals.\footnote[2]{For example, values of $\bar{k}$ 
	smaller or much smaller than $\sim 1$ imply that configurations with signals crossing Earth are rare or very rare.} 

\begin{figure}[t]
	\begin{center}
		\includegraphics[scale=0.38,clip=true]{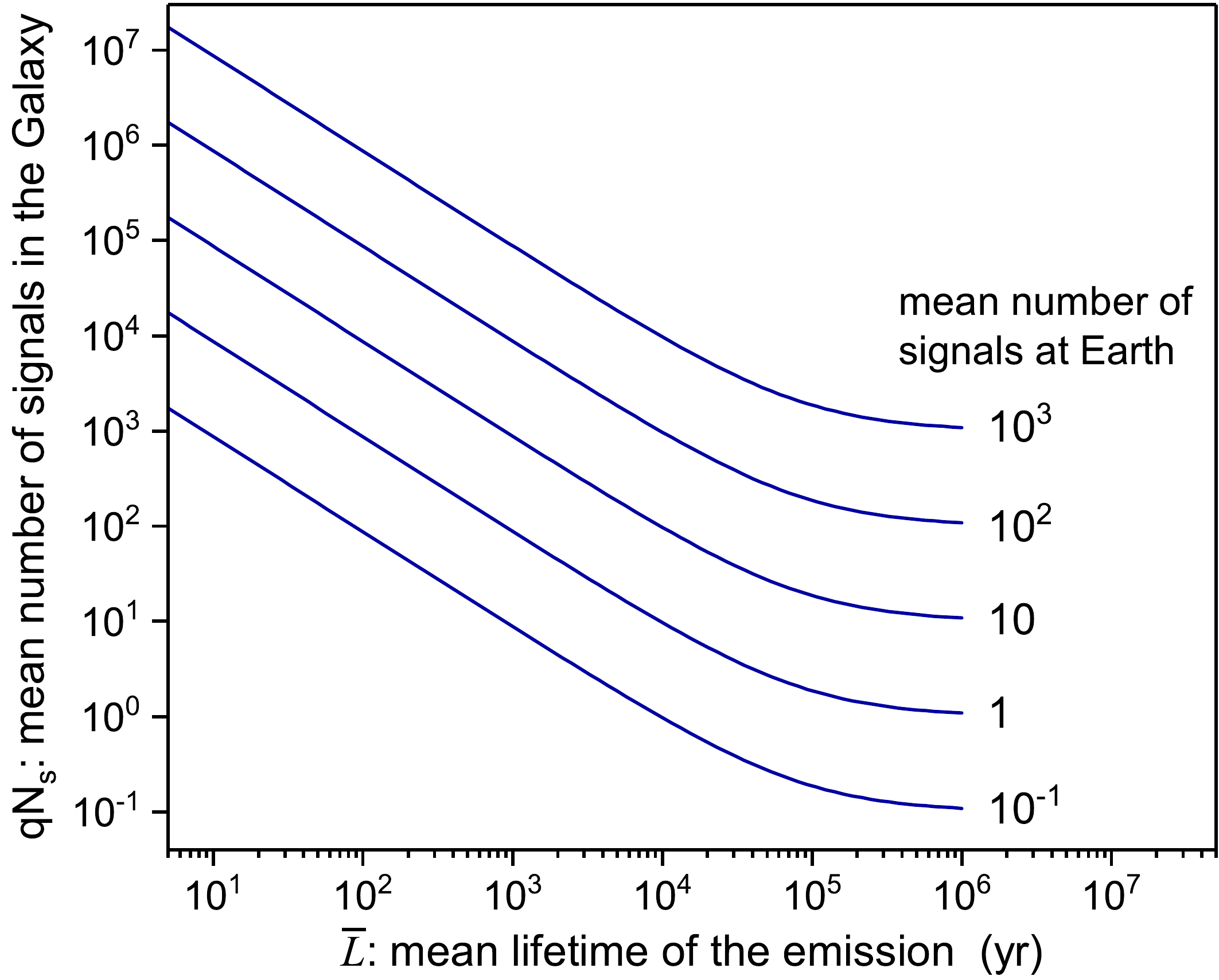}
		\caption{Relation between the average number of spherical shell signals present in the Galaxy, $qN_s$, and the mean signal longevity $\bar{L}$.
			The value of $qN_s$ for given $\bar{L}$ is determined by $\bar{k}$, the average number of signals crossing Earth, through
			$qN_s=\bar{k}(\bar{L}+t_M)/\bar{L}$, where $t_M$ is the age of the oldest signal which could possibly cross our planet. 
		}\label{fig2}
	\end{center}
\end{figure}

$\bar{k}$ is the quantity of main interest here for two reasons. First, Eq.~\ref{kbar} expresses the
two unknown quantities $\lambda$ and $q$ in terms of a single parameter, $\bar{k}$, which, as shown in the following,
can be in principle inferred by observations. 
As emphasized in Fig.~\ref{fig2}, knowledge of $\bar{k}$, or at least plausible upper or lower bounds, would also enable via Eq.~\ref{kbar} 
an estimate of the mean number $qN_s$ of shell signals occupying the Galaxy as a function of the mean signal longevity. For example,
$\bar{k}\sim 1$ implies that $qN_s$ can be as large as $\sim 1000$ if $\bar{L}\sim 100$ years is assumed, although in this case the vast 
majority of the signals do not cross the Earth.

Second, under the 
steady-state hypothesis, $\bar{k}$ coincides with the average number of emitters that are \emph{currently} radiating isotropic signals 
(SI Appendix, Section I). In particular, $\bar{k}$ can be shown to coincide with $\bar{L}/\tau$ \cite{Grimaldi2018}, where 
$\tau^{-1}$, the average birthrate of emitters, effectively incorporates the different probability factors appearing in the Drake 
equation \cite{Drake1961,Drake1965}. An informed estimate of $\bar{k}$ would bring therefore a valuable knowledge about the potential 
abundance of presently active emitters in the Galaxy.

\subsection{Signal detectability}
\label{sd}
In deriving Eq.~\ref{kbar}, we have considered possible galactic populations of isotropic signals without 
referring to their actual detectability by means of terrestrial, dedicated telescopes used in observational surveys. Even in the 
hypothesis that our planet lies in a region covered by the signals, their detection actually
depends on a number of factors such as the distance of the emitters, their radiated power, the wavelength of the 
signals, the minimum sensitivity of the detectors, and the search strategy.

To illustrate how these factors influence the detectability of extraterrestrial signals, we consider here the case in which a 
SETI search is designed to scan the entire sky for radio signals within a given range of frequencies. 
Contrary to targeted searches, in which a discrete set of target stars is selected, an all-sky survey covers in principle 
all directions of the sky. In this case the search space is
a sphere centered at Earth of radius specified by the radiated power of the emitter and by the detector sensitivity.
To see this, we assume that an emitter at $\vec{r}$ that transmits within a given range of radio frequencies has 
intrinsic luminosity $\textsf{L}$ (not to be confused with $L$, the signal longevity) and that in the same frequency range 
the detector (i.e., the radiotelescope) has a minimum detectable flux denoted $S_\textrm{min}$. Since the 
flux received by the detector is inversely proportional to the square 
of the distance from the source, the emitter is instrumentally detectable as long as its distance from the Earth is such that
\begin{equation}
\label{cond2}
\textsf{L}\geq 4\pi\vert \vec{r}-\vec{r}_0\vert^2S_\textrm{min}.
\end{equation}
The detection of an isotropic signal requires therefore that the conditions \ref{cond1} and \ref{cond2} must be simultaneously fulfilled. 
The resulting detection probability amounts to multiply $f_{R,\Delta}(\vec{r}-\vec{r}_o)$ in Eq.~\ref{p1} by
$\theta(R_\textsf{L}-\vert\vec{r}-\vec{r}_o\vert)$,
where $\theta(x)=1$ if $x\geq 0$ and $\theta(x)=0$ if $x<0$, and 
\begin{equation}
\label{RL}
R_\textsf{L}=\sqrt{\frac{\textsf{L}}{4\pi S_\textrm{min}}}
\end{equation}
is the distance beyond which an emitter with intrinsic luminosity $\textsf{L}$ is instrumentally undetectable. After the integrals
over $L$ and $t$ in Eq.~\ref{p1} are performed under the steady-state condition, the detection probability of a single signal reduces to:
\begin{equation}
\label{p2} p'=\frac{q}{N_s}\lambda\int\!d\vec{r}\rho_s(\vec{r})\theta(R_\textsf{L}-\vert\vec{r}-\vec{r}_o\vert),
\end{equation}
from which we recover Eq.~\ref{lambda} by choosing values of $\textsf{L}/S_\textrm{min}$ large enough to make 
$R_\textsf{L}$ bigger than $R_M$, the maximum distance of an emitter from Earth. 

Although $S_\textrm{min}$ is a known parameter that depends on the instrumental
characteristics of the detector, the intrinsic luminosity of the emitter, $\textsf{L}$, is an unknown quantity, which we treat 
probabilistically by introducing a PDF of the luminosity (commonly denoted luminosity function), 
$g(\textsf{L})$, independent of the duration of the emission process. We replace therefore
the detection probability given in Eq.~\ref{p2} by $p'=q\lambda\pi_o(R_{\textsf{L}^*})$, where
\begin{equation}
\label{pio}\pi_o(R_{\textsf{L}^*})=\frac{1}{N_s}\int_0^{\textsf{L}^*}\! d\textsf{L}g(\textsf{L})
\int\! d\vec{r}\rho_s(\vec{r})\theta(R_\textsf{L}-\vert\vec{r}-\vec{r}_o\vert)
\end{equation} 
is the luminosity detection probability. Although not strictly necessary, we have assumed in Eq.~\ref{pio}
that $g(\textsf{L})$ vanishes for luminosities larger than a maximum value, $\textsf{L}^*$. In this way, Eq.~\ref{pio} implies that
an emitter that is outside a sphere centered on Earth and of radius 
\begin{equation}
\label{RLstar}
R_{\textsf{L}^*}=\sqrt{\frac{\textsf{L}^*}{4\pi S_\textrm{min}}}
\end{equation}
is instrumentally undetectable, even if the emitted shell intersects the Earth. Note that since we take $\rho_s(\mathbf{r})$ 
to be approximately disk-like, the luminosity detection probability resulting from a SETI survey of the sky around
the galactic plane, instead of an all-sky survey, is not expected to differ significantly from Eq.~\ref{pio}.

Given $p'=q\lambda\pi_o$, where $\pi_o$ is a short-hand notation for $\pi_o(R_{\textsf{L}^*})$, and assuming that the emitters
have the same luminosity function, the probability that a telescope involved in the all-sky survey detects 
exactly $k=0,\, 1,\, 2\,\ldots,\, N_s$ spherical shell signals reduces to a binomial distribution: 
\begin{equation}
\label{binomial}
p(k\vert\pi_o)=\binom{N_s}{k}p'^k(1-p')^{N_s-k}.
\end{equation}
The average number of signals that can be detected by the survey is
therefore $\sum_{k=0}^{N_s} kp(k\vert\pi_o)=p'N_s=\pi_o\bar{k}$, where $\bar{k}$ is the mean number of signals at Earth given in Eq.~\ref{kbar}. 
Finally, noting that the value of $N_s$ inferred from the analysis of  the data from the Kepler space telescope is in the order of tens 
of billions \cite{Petigura2013}, Eq.~\ref{binomial} can be conveniently approximated by a Poisson distribution as long 
as $k$ and $\pi_o\bar{k}$ are much smaller than $N_s\approx 10^{10}$. We write therefore:
\begin{equation}
\label{poisson2}
p(k\vert\pi_o)=\frac{(\pi_o\bar{k})^k}{k!}e^{-\pi_o\bar{k}},
\end{equation}
which completes the definition of our model. In the following Bayesian analysis, we will use Eq.~\ref{poisson2} to derive the
likelihood functions corresponding to possible outcomes of a SETI search.

\section{Bayesian analysis}
\label{bayesanalysis}

\begin{figure*}[t]
	\begin{center}
		\includegraphics[scale=0.47,clip=true]{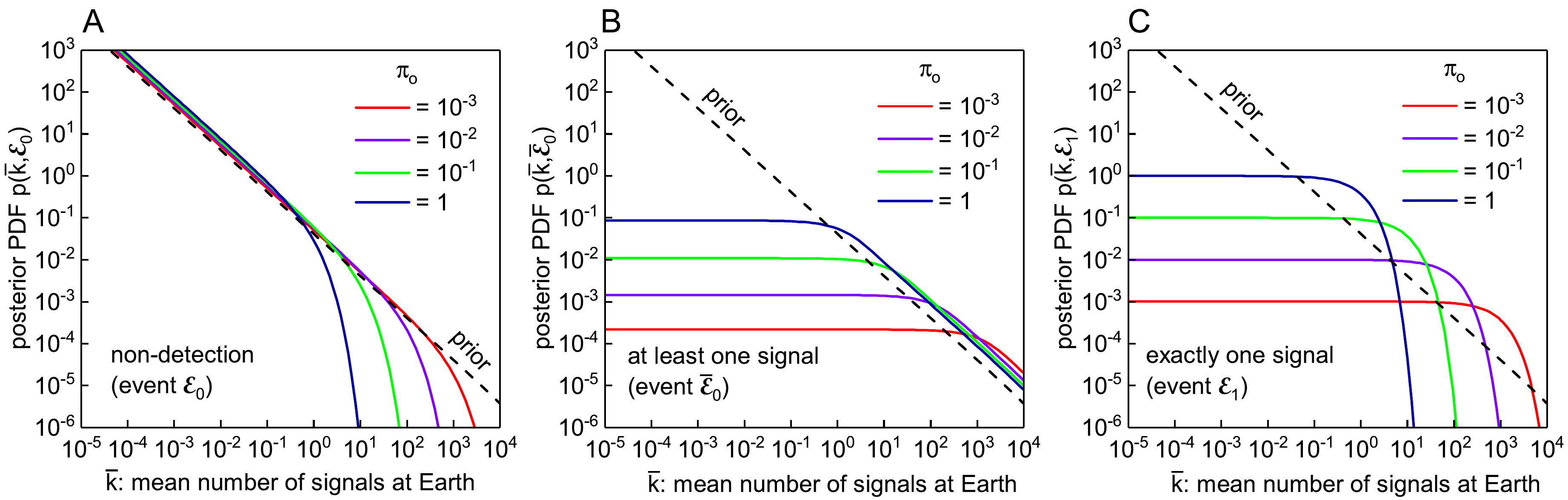}
		\caption{Probability distribution functions (PDFs) of the mean number of shell signals crossing Earth, $\bar{k}$, for different values of 
			the luminosity detection probability $\pi_o$. 
			The dashed and solid lines represent respectively the prior and posterior PDFs resulting from A: no signal detection (event $\mathcal{E}_0$), 
			B: at least one detectable signal (event $\overline{\mathcal{E}}_0$), C: exactly one detectable signal (event $\mathcal{E}_1$). 
			In A and B the posterior PDFs are smaller than the prior when, respectively, $\bar{k}>1/\pi_o$ and $\bar{k}<1/\pi_o$, whereas in C the weight
			of the posterior PDF is concentrated mostly around $\bar{k}\sim 1/\pi_o$. 
		}\label{fig3}
	\end{center}
\end{figure*}

Bayes' theorem provides a recipe for updating an initial hypothesis about the probability of occurrence of an event in 
response to new evidence \cite{Trotta2008}. Here, we take the initial hypothesis that the Earth intersects with a prior probability
distribution $p(\bar{k})$ 
an average number $\bar{k}\geq 0$ of signals emitted from communicating civilizations in the Galaxy, regardless 
of whether we detect them or not. 

Let us suppose that new evidence on the number of
detected signals (evidence $\mathcal{E}$) emerges from the acquisition of new data in a SETI survey.
Bayes' theorem states that the posterior probability that there are in average $\bar{k}$ signals intercepting
our planet taking into account the evidence $\mathcal{E}$ is:
\begin{equation}
\label{bayes1}
p(\bar{k}\vert\mathcal{E})=\frac{p(\mathcal{E}\vert\bar{k})p(\bar{k})}{p(\mathcal{E})},
\end{equation}
where $p(\mathcal{E})=\int\!d\bar{k}p(\mathcal{E}\vert\bar{k})p(\bar{k})$, the marginal likelihood of $\mathcal{E}$,
is a normalization constant 
and $p(\mathcal{E}\vert\bar{k})$ is the likelihood function defined as the conditional probability that the event 
$\mathcal{E}$ occurs given the initial hypothesis about $\bar{k}$.

\subsection{Likelihood terms}
\label{likes}
For the sake of simplicity, we shall not discuss here the occurrence of false positive or false negative 
results from an all-sky SETI survey, and we consider the only two possible outcomes, that is, a negative result for signal 
detection or a positive evidence for the existence of communicating civilizations represented by the detection of one signal.

In the first case we assume that an observational campaign as the one described in Sec.~\ref{model} has detected no signals
within the entire sky (and within a depth set by $R_{\textsf{L}^*}$). Let 
$\mathcal{E}_0$ denote this evidence. The corresponding likelihood function, $p(\mathcal{E}_0\vert\bar{k})$,  is obtained by
setting $k=0$ in Eq.~\ref{poisson2},
\begin{equation}
\label{likeE0}
p\left(\mathcal{E}_0\vert\bar{k}\right)=e^{-\pi_o\bar{k}}.
\end{equation}
Quite intuitively, Eq.~\ref{likeE0} shows that for $\pi_o \neq 0$ the probability of $\mathcal{E}_0$ occurring decays exponentially 
with $\bar{k}$, implying that values of $\bar{k}$ much larger than $1/\pi_o$ can be ruled out by the non-detection of signals.

\begin{figure*}[t]
	\begin{center}
		\includegraphics[scale=0.47,clip=true]{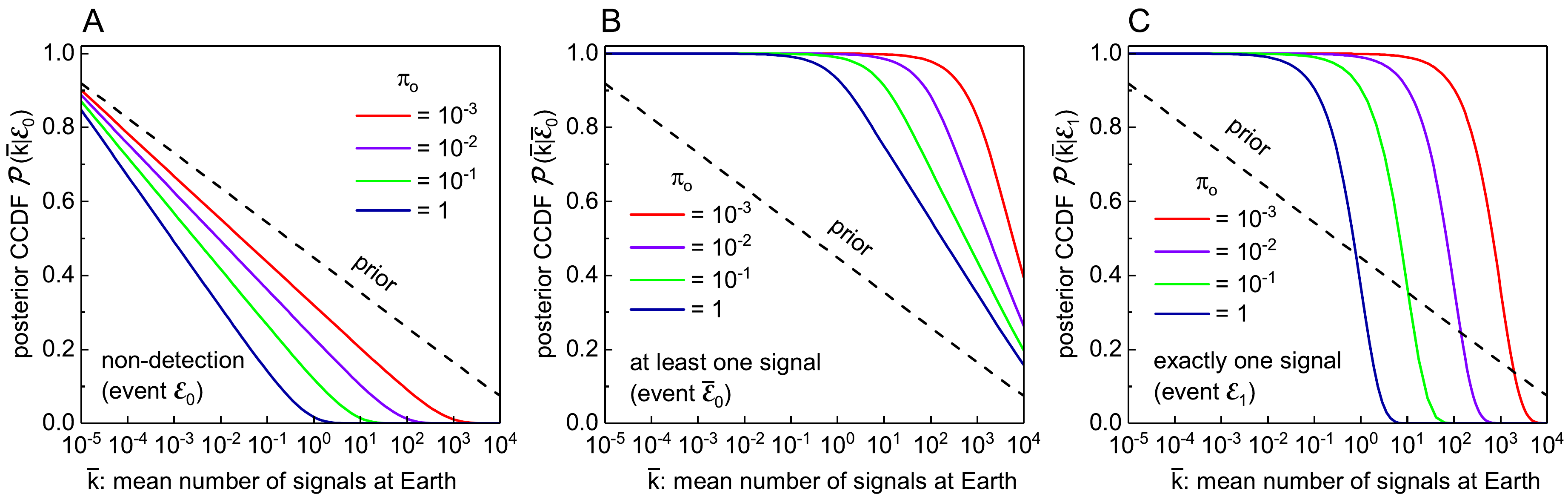}
		\caption{Complementary cumulative distribution functions (CCDFs) of $\bar{k}$ for different values of the luminosity detection probability $\pi_o$. 
			Each curve gives the probability that the mean number of signals intersecting Earth's orbit is larger than $\bar{k}$.
			The dashed and solid lines represent respectively the prior and posterior CCDFs resulting from A: no signal detection (event $\mathcal{E}_0$), 
			B: at least one detectable signal (event $\overline{\mathcal{E}}_0$), C: exactly one detectable signal (event $\mathcal{E}_1$). 
			In A the posterior CCDF becomes progressively smaller than the prior as $\pi_o$ increases, and it vanishes exponentially for $\bar{k}>1/\pi_o$.
			in B and C the posterior CCDFs deviate more form the prior when $\pi_o$ is smaller. For $\pi_o=10^{-3}$ the posterior probability that there
			are more than $100$ signals intercepting the Earth is larger than $95$ \%.
		}\label{fig4}
	\end{center}
\end{figure*}

In considering the case that an all-sky SETI survey detects a non-natural, extraterrestrial signal, we need to distinguish between two
possibilities depending on whether the gathered evidence can exclude or not the existence of detectable signals from other emitters
within $R_{\textsf{L}^*}$ (besides the one already detected). This distinction has to be made because the detection may occur
before the sky has been entirely swept out, not excluding therefore the possibility that there may be other detectable signals 
from emitters within $R_{\textsf{L}^*}$. In this case the evidence, denoted $\overline{\mathcal{E}}_0$, is that there is \emph{at least one} 
detectable signal emitted within a distance $R_{\textsf{L}^*}$. The associated likelihood is $p(\overline{\mathcal{E}}_0\vert\bar{k})=1-p(\mathcal{E}_0\vert\bar{k})$ 
because $\overline{\mathcal{E}}_0$ is the negation of $\mathcal{E}_0$, the event of non-detection considered above. Hence
\begin{equation}
\label{likenotE0}
p\left(\overline{\mathcal{E}}_0\vert\bar{k}\right)=1-e^{-\pi_o\bar{k}}.
\end{equation}
The second possibility is that the evidence, denoted $\mathcal{E}_1$, amounts to detect \emph{exactly one} emitter 
in the entire sky within a depth $R_{\textsf{L}^*}$, as it would be the case if no other 
signals have been detected upon the completion of the survey.  The likelihood term in this case is therefore given by 
Eq.~\ref{poisson2} with $k=1$: 
\begin{equation}
\label{likeE1}
p\left(\mathcal{E}_1\vert\bar{k}\right)=\pi_o\bar{k} e^{-\pi_o\bar{k}}.
\end{equation}
Since the likelihoods \ref{likenotE0} and ~\ref{likeE1} are significant when, respectively, $\pi_o\bar{k}\gtrsim 1$ and 
$\pi_o\bar{k}\sim 1$, large values of $\bar{k}$ have to be expected when $\pi_o$ is small.
In other terms, the smaller the fraction of the Galaxy in which a SETI survey is successful, the larger is the likely number of broadcasting 
emitters in the Milky Way.

\subsection{Prior distribution}
\label{priors}
To obtain the posterior probability $p(\bar{k}\vert\mathcal{E})$, $\mathcal{E}=\mathcal{E}_0$, $\mathcal{E}_1$, and $\overline{\mathcal{E}}_0$, 
we need to specify $p(\bar{k})$, the prior probability distribution of $\bar{k}$. Presently, we lack generally accepted arguments to 
estimating either the fraction $q$ of stars in the Galaxy that may harbor communicating civilizations, or the mean signal longevity $\bar{L}$. 
Possible values of $\bar{k}$ may therefore range from $\bar{k}=0$, as argued by proponents of the rare Earth hypothesis \cite{Ward2000}, to a 
significant fraction of $N_s$, in the most optimistic scenarios. 

A natural choice of $p(\bar{k})$, befitting our ignorance about even the scale of $\bar{k}$, would be taking
a prior PDF that is uniform in $\log(\bar{k})$, which corresponds to $p(\bar{k})\propto \bar{k}^{-1}$, to give equal 
weight to all orders of magnitude \cite{Trotta2008,Spiegel2012}.\footnote[3]{At first sight, a prior PDF uniform in $\bar{k}$ appears
	to reflect our state of ignorance. It is however an highly informative prior because it strongly favors large values of $\bar{k}$ \cite{Spiegel2012}.} 
Although the log-uniform prior appropriately expresses our state of ignorance, it fails to take into account 
that, after all, various past SETI surveys have been conducted since several decades \cite{Tarter2001}, with null results. 
Likewise, there have been no serendipitous detection of non-natural extraterrestrial signals since the invention of radio telescopes. 

To allow the prior PDF to reflect the so far lack of detection, we introduce a prior luminosity detection probability defined as
$\pi_o^\textrm{prior}=\pi_o(R_{\textsf{L}^*}^\textrm{prior})$, where $R_{\textsf{L}^*}^\textrm{prior}$, the prior observational radius, is 
representative of the distance accessible by past SETI surveys for given values of $\textsf{L}^*$ and frequency range.
The likelihood of non-detection, Eq.~\ref{likeE0}, immediately suggests that a natural way to inform the prior 
about past SETI negative results is to update $p(\bar{k})$ using Bayes' theorem, leading to $p(\bar{k})\propto\bar{k}^{-1}\exp(-\pi_o^\textrm{prior}\bar{k})$.
In so doing, we are simply adopting as prior the posterior PDF resulting from the non-detection of signals of 
past SETI initiatives. Finally, in order to make the prior distribution proper (i.e., normalizable) we introduce a lower cut-off in $\bar{k}$ 
so that $p(\bar{k})=0$ when $\bar{k}<\bar{k}_\textrm{min}$. The normalized prior distribution becomes therefore
\begin{equation}
\label{prior}
p(\bar{k})=\frac{\bar{k}^{-1}e^{-\displaystyle \pi_o^\textrm{prior}\bar{k}}}{E_1(\pi_o^\textrm{prior}\bar{k}_\textrm{min})}\theta(\bar{k}-\bar{k}_\textrm{min}),
\end{equation}
where $E_1(x)=\int_x^\infty\! dt\, e^{-t}/t$ is the exponential integral. The value of $\bar{k}_\textrm{min}$ can be chosen so as to satisfy some
appropriate criterion. Here we adopt the requirement that $\bar{k}_\textrm{min}$ gives the least informative prior probability 
that at least one signal from the entire Galaxy intercepts Earth's orbit, leading to $\bar{k}_\textrm{min}\simeq 0.14\pi_o^\textrm{prior}$ (SI Appendix, Section II).

\begin{figure*}[t]
	\begin{center}
		\includegraphics[scale=0.4,clip=true]{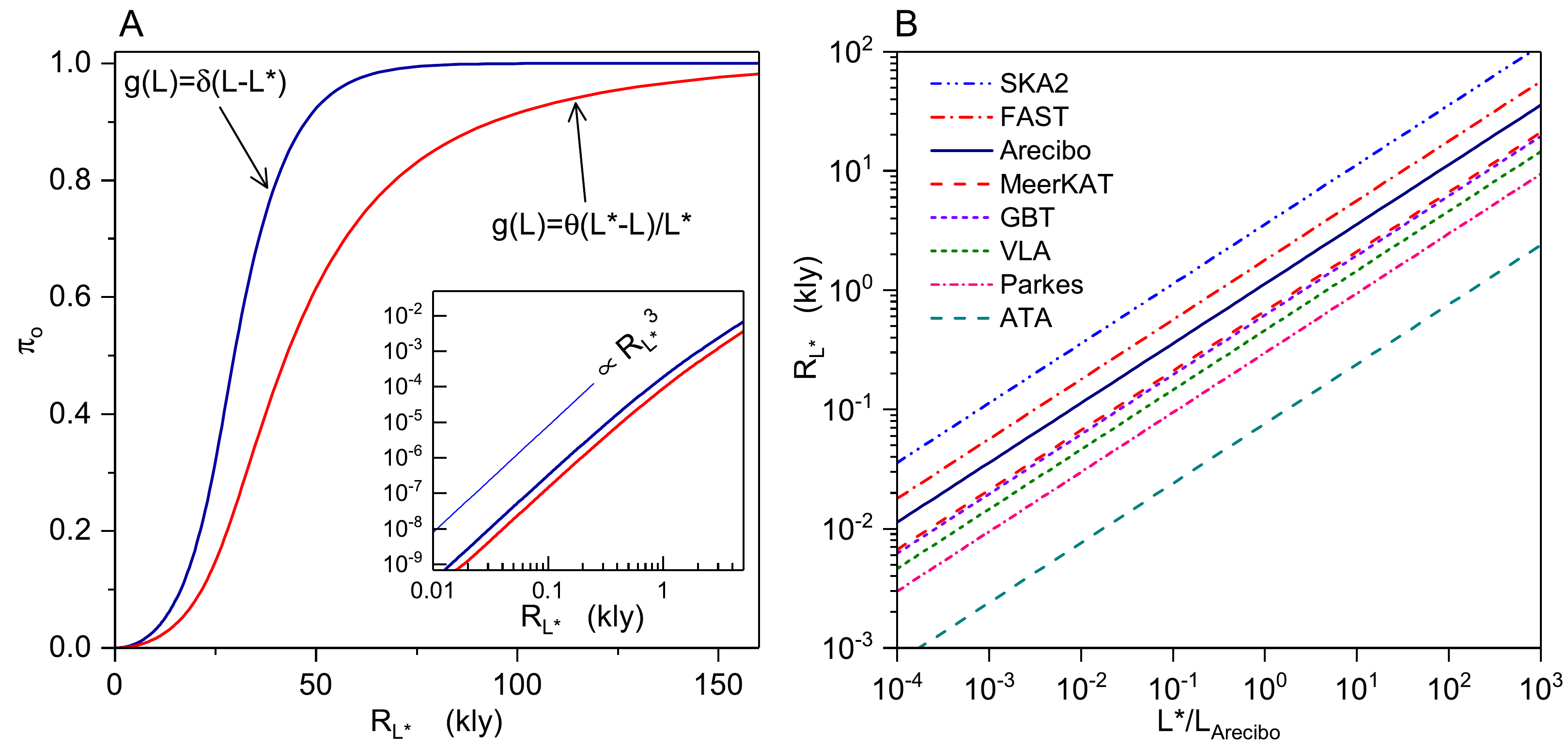}
		\caption{A: Probability $\pi_o$ that an emitter is within an observable sphere of radius $R_{\textsf{L}^*}$ for the cases in which the emitter
			luminosity function is either a single Dirac-delta peak centered at $\textsf{L}^*$ or a uniform distribution extending up to  $\textsf{L}^*$.
			The inset shows that within the galactic neighborhood
			($R_{\textsf{L}^*}\lesssim 1$ kly) $\pi_o$ scales as $R_{\textsf{L}^*}^{\,\,\, 3}$. B: Radius of the observable sphere for 
			several telescopes as a function
			of the intrinsic luminosity of an emitter in units of $\textsf{L}_\textrm{Arecibo}=2\times 10^{13}$ W, the equivalent isotropic radiated 
			power of the Arecibo radar. The values of the minimum detectable flux ($S_\textrm{min}$) are reported in 
			Table~\ref{table1} for each telescope. Note that from Eq.~\ref{RLstar} $R_{\textsf{L}^*}$ scales as $\sqrt{\textsf{L}^*}$.
		}\label{fig5}
	\end{center}
\end{figure*}

\subsection{Posterior probabilities}
\label{post}

The normalized product of the prior probability distribution \ref{prior} with each of the three likelihood functions \ref{likeE0}, 
\ref{likenotE0}, and \ref{likeE1} gives the respective PDFs resulting from the events $\mathcal{E}_0$ (non-detection),
$\overline{\mathcal{E}}_0$ (at least one detection), and $\mathcal{E}_1$ (exactly one detection). 
To keep the analysis as general as possible, for the moment we treat the luminosity detection probability, $\pi_o$, as an independent variable 
ranging between $0$ and $1$. We shall restore its full dependence upon the observational radius $R_{\textsf{L}^*}$ in the section dedicated to 
the discussion of present and future SETI surveys. 
For illustrative purposes, we also take $\pi_o^\textrm{prior}$ constant and equal to $10^{-5}$, a value not far from our subsequent estimates
of the prior luminosity detection probability.

Figure~\ref{fig3} compares the PDFs of $\bar{k}$ (solid lines), calculated  for different values of $\pi_o$,
with the prior distribution (dashed lines). There are three relevant features worth to be stressed. First, all three posteriors are manifestly 
driven by their respective data ($\mathcal{E}_0$, $\overline{\mathcal{E}}_0$, and $\mathcal{E}_1$) and not the prior, confirming that the 
latter is fairly non-informative. Second, the posterior PDF resulting from the event of non-detection,
Fig.~\ref{fig3}A, converges smoothly to the prior PDF as $\pi_o\rightarrow 0$, and deviates most from it
when $\bar{k}>1/\pi_o$, as anticipated by the likelihood term \ref{likeE0}. A substantial effect of the datum (that is, the event of non-detection) is expected therefore
only for $\pi_o$ significantly larger than $\pi_o^\textrm{prior}$. Finally, the third significant result is that
the posterior PDFs resulting from the detection of a signal, Fig.~\ref{fig3}B and \ref{fig3}C, do not converge to the prior for 
$\pi_o\rightarrow 0$. In this limit, the corresponding
likelihood terms \ref{likenotE0} and \ref{likeE1} are proportional to $\bar{k}$, which cancels the factor $\bar{k}^{-1}$
of the prior. Consequently, for $\bar{k}\lesssim 1/\pi_o$ the posterior PDFs $p(\bar{k}\vert\overline{\mathcal{E}}_0)$ and
$p(\bar{k}\vert\mathcal{E}_1)$ are approximately constant, and get progressively small as $\pi_o$ diminishes. 
This has the net effect of shifting the weight of $p(\bar{k}\vert\overline{\mathcal{E}}_0)$,  Fig.~\ref{fig3}C, to
$\bar{k}>1/\pi_o$ and, due to the cutoff at $\bar{k}>1/\pi_o$ in the likelihood term associated to $\mathcal{E}_1$, of concentrating
the weight of $p(\bar{k}\vert\mathcal{E}_1)$ in the region around $\bar{k}\sim 1/\pi_o$, Fig.~\ref{fig3}C. Therefore, in case of detection,
the smaller the value of $\pi_o$ (or, equivalently, the smaller the observational radius $R_{\textsf{L}^*}$) the larger is the probability 
that the Earth intersects many shell signals other than the one already detected.

By integrating the posterior PDFs from $\bar{k}$ to $\infty$, we calculate the complementary cumulative distribution functions
(CCDFs), which give the updated probabilities that the mean number of shells intersecting Earth, transmitted from the entire Galaxy,
is larger than $\bar{k}$. For each event $\mathcal{E}_0$, $\overline{\mathcal{E}}_0$, and $\mathcal{E}_1$, the CCDFs are given respectively by:
\begin{align}
\label{CCDFE0}
\mathcal{P}(\bar{k}\vert\mathcal{E}_0)&=\int_{\bar{k}}^\infty\!d\bar{k}'p(\bar{k}'\vert\mathcal{E}_0)=
\frac{E_1[(\pi_o+\pi_o^\textrm{prior})\bar{k}]}{E_1[(\pi_o+\pi_o^\textrm{prior})\bar{k}_\textrm{min}]}, \\
\label{CCDFnotE0}
\mathcal{P}(\bar{k}\vert\overline{\mathcal{E}}_0)&=\int_{\bar{k}}^\infty\!d\bar{k}'p(\bar{k}'\vert\overline{\mathcal{E}}_0)\nonumber \\
&=\frac{E_1(\pi_o^\textrm{prior}\bar{k})-E_1[(\pi_o+\pi_o^\textrm{prior})\bar{k}]}
{E_1(\pi_o^\textrm{prior}\bar{k}_\textrm{min})-E_1[(\pi_o+\pi_o^\textrm{prior})\bar{k}_\textrm{min}]},\\
\label{CCDFE1}
\mathcal{P}(\bar{k}\vert\mathcal{E}_1)&=\int_{\bar{k}}^\infty\!d\bar{k}'p(\bar{k}'\vert\mathcal{E}_1)=
e^{-\displaystyle (\pi_o+\pi_o^\textrm{prior})(\bar{k}-\bar{k}_\textrm{min})}.
\end{align}
Figure \ref{fig4} shows Eqs.~\ref{CCDFE0}-\ref{CCDFE1} (solid lines) for the same values of $\pi_o$ of Fig.~\ref{fig3}. 
The prior CCDF (dashed lines) is obtained by setting $\pi_o=0$ in Eq.~\ref{CCDFE0}. 
As anticipated by the analysis of the PDFs, the posterior CCDFs resulting from the non-detection or a detection of a signal differ 
significantly from each other. While the response to $\mathcal{E}_0$, Fig.~\ref{fig4}A, becomes progressively smaller 
than the prior as $\pi_o$ increases, and getting negligibly small for $\bar{k}>1/\pi_o$, the posterior CCDFs resulting from the detection
of a signal (either event $\overline{\mathcal{E}}_0$ or $\mathcal{E}_1$) depart abruptly from the prior CCDF as soon as $\pi_o\neq 0$, 
as shown in Figs.~\ref{fig4}B and \ref{fig4}C.
In particular, for a relatively small value of $\pi_o$ (say for example $\sim 10^{-3}$) the detection of at least one signal
(event $\overline{\mathcal{E}_0}$) implies that the posterior probability that the Earth intersects typically more than $\bar{k}\sim 1/\pi_o\sim 10^3$ 
signals exceeds $\sim 80$\%. This probability drops to about $35$\% if exactly one signal is detected in the entire sky for the same value of $\pi_o$ (event $\mathcal{E}_1$, Fig.~\ref{fig3}c). Since $\mathcal{P}(\bar{k}\vert\overline{\mathcal{E}_0})$ and $\mathcal{P}(\bar{k}\vert\mathcal{E}_1)$ 
represent respectively an upper and lower limit for the posterior probability in the case of detection, the probability that 
there are in average more than $10^3$ galactic signals crossing the Earth is therefore comprised between $\sim 80$\% and $\sim 35$\% in this example.

\begin{table}[t]
	\caption{System equivalent flux density, $S_\textrm{sys}$, and corresponding sensitivity, $S_\textrm{min}$, for the Allen Telescope Array (ATA) \cite{Enriquez2017,Harp2016},
		the Parkes telescope \cite{Enriquez2017}, the Jansky Very Large Array telescope (VLA) \cite{Enriquez2017,Gray2017}, 
		the Green Bank Telescope (GBT) \cite{Enriquez2017}, the Meer Karoo
		Array Telescope (MeerKAT) \cite{Jonas2018}, the Arecibo telescope \cite{Foster2018}, the Five hundred meter Aperture Spherical Telescope (FAST) \cite{Nan2011}, 
		and the Phase 2 of the Square Kilometre Array (SKA2) \cite{Lazio2013}. The values of $S_\textrm{min}$ are calculated from Eq.~\ref{Smin} 
		assuming $m=15$, $\Delta\nu=0.5$ Hz, and $t=10$ min.}
	\label{table1}
	\begin{ruledtabular}
	\begin{tabular}{ccc}
		telescope & $S_\textrm{sys}$ (Jy) & $S_\textrm{min}$ ($10^{-26}$ W/m$^2$)\\
		\hline
		ATA & $664$\footnotemark[1] & $287$ \\
		Parkes & $43$ & $18.6$ \\
		VLA & $18$ & $7.8$ \\
		GBT & $10$ & $4.3$ \\
		MeerKAT & $8.6$ & $3.7$ \\
		Arecibo & $3$ & $1.3$ \\
		FAST & $1.2$ & $0.5$ \\
		SKA & $0.3$\footnotemark[2] & $0.13$ \\
	\end{tabular}
\end{ruledtabular}	
\footnotetext[1]{SEFD for $27$ antennas}
\footnotetext[2]{Estimate of the goal value the SEFD (frequency range $\sim 1$-$2$ GHz) targeted upon completion of the phase 2 of the 
	SKA telescope \cite{Lazio2013}. See \texttt{https://astronomers.skatelescope.org/documents} for further documentation on phase 1, phase 2, and precursors of SKA.}

\end{table}

\begin{figure}[t]
	\begin{center}
		\includegraphics[scale=0.55,clip=true]{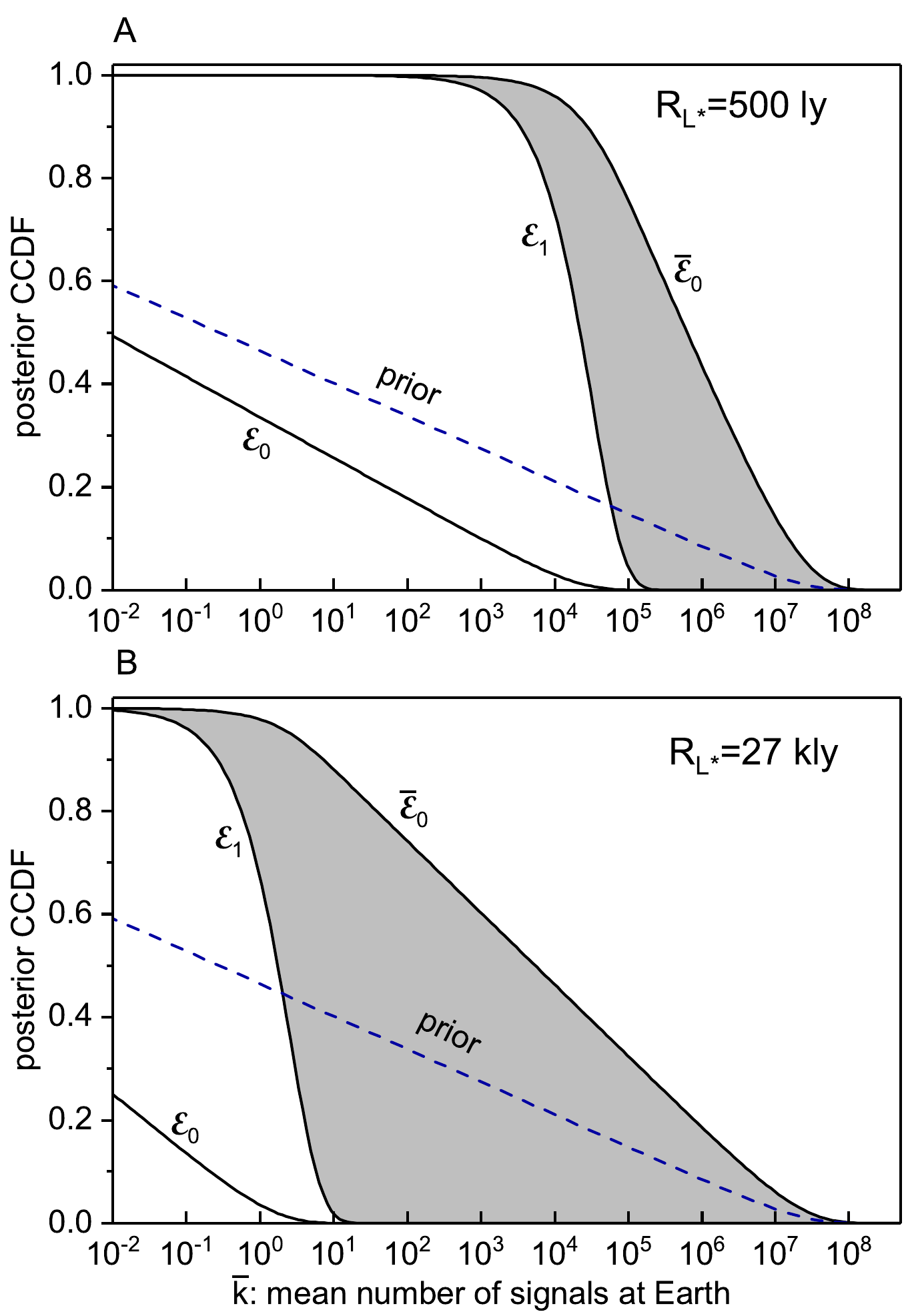}
		\caption{Posterior CCDFs giving the probabilities that the mean number of signals crossing Earth is larger than $\bar{k}$. The solid curves refer
			to the CCDFs resulting from the events of non-detection ($\mathcal{E}_0$), at least one detectable signal ($\overline{\mathcal{E}}_0$), and exactly 
			one detectable signal ($\mathcal{E}_1$) within $500$ ly from Earth (A), corresponding to an observational radius containing 
			about one million nearby stars targeted by the ``Breakthrough Listen" project, or within $27$ kly from Earth (B), that is the distance to the galactic center. 
			The dashed line denotes the prior CCDF calculated as described in the text.
			The results are computed by adopting a delta-Dirac luminosity function for the emitters centered at $\textsf{L}^*=\textsf{L}_\textrm{Arecibo}$, where
			$\textsf{L}_\textrm{Arecibo}=2\times 10^{13}$ W is the equivalent isotropic radiated power (EIRP) of the Arecibo radar.}\label{fig6}
	\end{center}
\end{figure}

\section{Bayesian analysis applied to existing and upcoming SETI detectors}
\label{bayanalysis}

We now apply our Bayesian formalism by considering existing and planned radiotelescopes to infer the posterior probabilities following the 
potential occurrence of events $\mathcal{E}_0$, $\overline{\mathcal{E}}_0$, and $\mathcal{E}_1$ in a SETI survey.
The quantity governing the response to these events is $\pi_o(R_{\textsf{L}^*})$, the probability that an emitter is within
a distance $R_{\textsf{L}^*}$ from the Earth. According to Eq.~\ref{pio}, this quantity depends on the number 
distribution of stars, $\rho_s(\vec{r})$, and the luminosity function of the emitters, $g(\textsf{L})$. The latter 
identifies also the observational radius, $R_{\textsf{L}^*}$, once a specific telescope sensitivity is assigned. 

\subsection{Number density of stars}
We take the number density function $\rho_s(\vec{r})$ to have a cylindrical symmetry of the form:
\begin{equation}
\label{ghz}
\rho_s(\vec{r})=\rho_0 e^{-\displaystyle r/r_s}e^{-\displaystyle \vert z\vert/z_s}
\end{equation}
where $r$ is the radial distance from the galactic center, $z$ is the height from the galactic plane, and $\rho_0$ is a normalization
factor ensuring that $\int\!d\vec{r}\rho_s(\vec{r})=N_s$. 
We assume that the emitters are potentially located in the thin disk of the Milky Way, whose star distribution follows approximately
Eq.~\ref{ghz} with  $r_s=8.15$ kly and $z_s=0.52$ kly \cite{Misiriotis2006}.
The resulting $\rho_s(\vec{r})/N_s$ gives a probability over $99$ \% of finding a star at
a distance of $60$ kly from the galactic center.\footnote[4]{We have considered 
	also the possibility that the distribution of stars that can potentially harbor life
	has an annular shape, as in the galactic habitable zone proposed in Ref.~\cite{Lineweaver2004} (SI Appendix, Section 3).}

\subsection{Luminosity function}
Since we ignore what a plausible PDF of $\textsf{L}$ looks like, and whether it even exists, modeling the luminosity function $g(\textsf{L})$ unavoidably requires
making some assumptions. Previously, a power-law distribution of the form $g(\textsf{L})\propto L^{-\alpha}$ has been proposed as a vehicle
to assess, through the choice of the exponent $\alpha$, the interplay between the proximity of a detectable emitter and its spectral 
density \cite{Gulkis1985,Shostak2000}. Here, we limit our analysis to the effect on the detection probability of the spread of the luminosity distribution 
by considering some limiting forms of $g(\textsf{L})$. To this end, we take $g(\textsf{L})$ to be given  
either by a single Dirac-delta peak centered at some characteristic luminosity $\textsf{L}^*$, 
$g(\textsf{L})=\delta(\textsf{L}-\textsf{L}^*)$, or by a uniform distribution
ranging from $\textsf{L}=0$ up to $\textsf{L}^*$: $g(\textsf{L})=\theta(\textsf{L}^*-\textsf{L})/\textsf{L}^*$. 
In either case, the dependence of $\pi_o$ on the width of the luminosity function can be conveniently expressed in terms of the maximum detectable
distance, $R_{\textsf{L}^*}$, Eq.~\ref{RLstar}, as illustrated in Fig.~\ref{fig5}A. While at distances of about $R_\textrm{M}\sim 90$ kly or 
larger, $\pi_o$ saturates to one due to the finite size of the Galaxy, in the galactic neighborhood ($R_{\textsf{L}^*}\lesssim 1$ kly)
the function $\pi_o$ is smaller than about $10^{-3}$ and it scales as $R_{\textsf{L}^*}^{\,\,3}$, inset of Fig.~\ref{fig5}A.

\subsection{Observational radius}
To determine the observational radius $R_{\textsf{L}^*}$, Eq.~\ref{RLstar}, of an all-sky SETI survey, we must specify the characteristic luminosity of 
the emitters, $\textsf{L}^*$, and the minimum detectable flux, $S_\textrm{min}$. The latter quantity is determined by the characteristics of the telescope
used in the SETI search and the intrinsic bandwidth of the transmitted signal. Here, we consider the case of a signal
bandwidth narrower than the spectral resolution of the telescope, which reduces $S_\textrm{min}$ to \cite{Enriquez2017,Gray2017}:
\begin{equation}
\label{Smin}
S_\textrm{min}=m S_\textrm{sys}\sqrt{\frac{\Delta\nu}{t}},
\end{equation}
where $m$ is the desired signal-to-noise ratio, $t$ is the integration time in seconds, $\Delta\nu$ is the receiver channel bandwidth (expressed in Hz),
and $S_\textrm{sys}$ is the system equivalent flux density (SEFD), which depends on the system temperature of the receiver and on the effective collecting
area of the telescope. Table~\ref{table1} lists the values of $S_\textrm{sys}$ in Jy ($1$ Jy $=10^{-26}$ Wm$^{-2}$Hz$^{-1}$)
of a few existing and planned facilities \cite{Enriquez2017,Siemion2015,Harp2016,Gray2017,Nan2011,Jonas2018,Foster2018} and the corresponding 
$S_\textrm{min}$ in Wm$^{-2}$ calculated for $m=15$, $\Delta\nu=0.5$ Hz, and $t=600$ s. In Table~\ref{table1}, the
SEFD values of ATA, Parkes, VLA, GBT, and Arecibo refer to past targeted searches for non-natural radio signals of frequencies comprised 
between $\sim 1$ and $\sim 2$ GHz, while those attributed to MeerKAT, FAST and SKA2 are only indicative, as these
telescopes are either not yet fully operational (MeerKAT and FAST) or still in the study phase (SKA2). 

Figure \ref{fig5}B shows the values of $R_{L^*}$ that the telescopes enlisted in Table \ref{table1},
or other facilities of comparable sensitivity, could access if they were employed in an all-sky search for signals within $\sim 1$-$2$ GHz. 
From Figs.~\ref{fig5}A and \ref{fig5}B we see that for $\textsf{L}^*/\textsf{L}_\textrm{Arecibo}\lesssim 0.1$, where  
$\textsf{L}_\textrm{Arecibo}=2\times 10^{13}$ W is the equivalent isotropic radiated power (EIRP) emitted by the Arecibo 
radar,\footnote[5]{Although this value of the EIRP refers to the Arecibo radar transmitting at a 
	frequency of $2.38$ GHz, which is outside the observational frequency coverage considered here (about $1$-$2$ GHz), 
	we nonetheless adopt $\textsf{L}_\textrm{Arecibo}=2\times 10^{13}$ W as a useful term of comparison, as the Arecibo radar
	is the most powerful radio transmitter on Earth.} even the most sensitive 
receiver (the planned phase 2 of the SKA telescope) can probe distances only up to about $1$ kly, where the probability $\pi_o$
is small, while $\textsf{L}^*/\textsf{L}_\textrm{Arecibo}\gtrsim 100$ has to be assumed to make $\pi_o$ significant.

\begin{figure*}[t]
	\begin{center}
		\includegraphics[scale=0.7,clip=true]{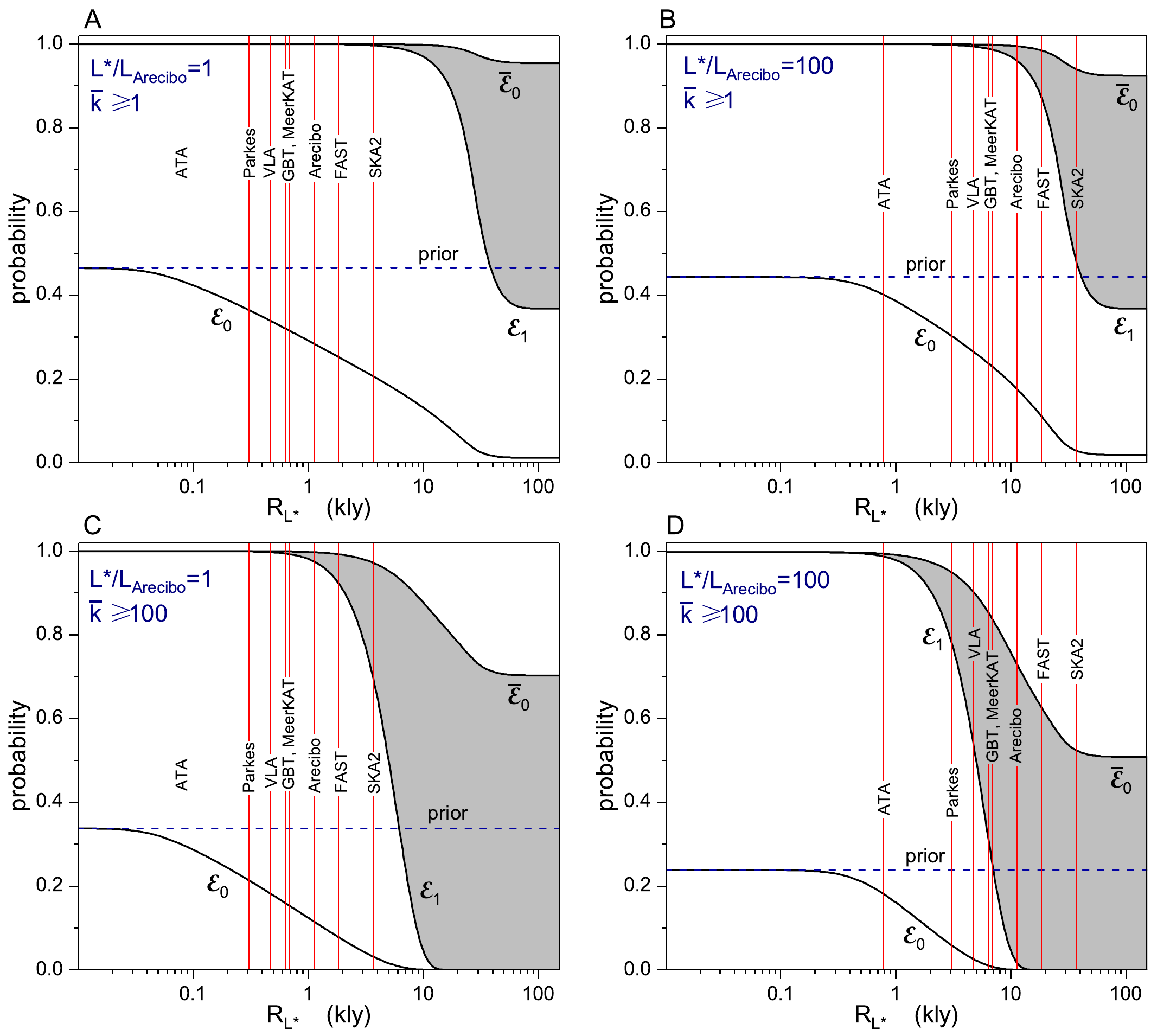}
		\caption{Posterior probability that $\bar{k}\geq 1$ (top row) and $\bar{k}\geq 100$ (bottom row) for emitters with characteristic luminosity 
			$\textsf{L}^*/\textsf{L}_\textrm{Arecibo}=1$  (left column) and $\textsf{L}^*/\textsf{L}_\textrm{Arecibo}=100$ (right column), where
			$\textsf{L}_\textrm{Arecibo}=2\times 10^{13}$ W is the EIRP of the Arecibo radar. 
			Dashed lines denote the prior probabilities, while the solid curves are posterior probabilities as a function of the observational radius $R_{\textsf{L}^*}$
			resulting from the events of non-detection ($\mathcal{E}_0$), at least one detectable signal ($\overline{\mathcal{E}}_0$), and exactly 
			one detectable signal ($\mathcal{E}_1$). The results have been obtained by assuming a Dirac-delta luminosity
			function centered at $\textsf{L}^*$.
			The red vertical lines indicate the values of $R_{\textsf{L}^*}$ that are accessible to the telescopes listed in Table~\ref{table1}.}
		\label{fig7}
	\end{center}
\end{figure*}

\subsection{Prior parameters}

To assign the probability $\pi_o^\textrm{prior}$ that has to be plugged into the prior PDF \ref{prior}, we have to estimate the observational radii within which
no emitters have been detected so far, assuming given EIRP values of the emitters. 
The best strategy is probably to adopt a value of $S_\textrm{min}$, denoted $S_\textrm{min}^\textrm{prior}$, that represents an effective detection
threshold combining previous SETI surveys. Depending on the detectors employed and their location, previous sky surveys covered different extended regions
of the sky with flux thresholds ranging from about $10^{-22}$ W/m$^2$ to about $10^{-24}$ W/m$^2$ in the frequency range $1$-$2$ GHz\cite{Tarter2001,Wolfe1981,Horowitz1993,Colomb1993,Stootman2000,Leigh2000,Bowyer2016}.
Here we adopt a conservative value of $S_\textrm{min}^\textrm{prior}=10^{-23}$ Wm$^{-2}$, 
which brings about a maximum detectable radius of $R_{\textsf{L}^*}^\textrm{prior}=0.0422\sqrt{\textsf{L}^*/\textsf{L}_\textrm{Arecibo}}$ kly.
In other terms, by adopting $S_\textrm{min}^\textrm{prior}=10^{-23}$ Wm$^{-2}$ we are ruling out the existence of detectable signals from emitters 
within $\sim 40$ ly from Earth that transmit within a frequency range $1$-$2$ GHz with an EIRP equivalent to that of the Arecibo radar.

Even for $\textsf{L}^*=100\textsf{L}_\textrm{Arecibo}$,  the chosen $R_{\textsf{L}^*}^\textrm{prior}$ is well within the distance at which the probability of 
detecting an emitter luminosity is small and follows a power law. For a Dirac-delta luminosity function we estimate 
$\pi_o^\textrm{prior}=\pi_o(R_{\textsf{L}^*}^\textrm{prior})\sim 2.6\times 10^{-8}(\textsf{L}^*/\textsf{L}_\textrm{Arecibo})^{3/2}$
(SI Appendix, Section III). The corresponding value of $\bar{k}_\textrm{min}$ follows from $\bar{k}_\textrm{min}\simeq 0.14\pi_o^\textrm{prior}$, as 
discussed above.

\subsection{Posteriors}
Figure \ref{fig6} shows the posterior CCDFs of $\bar{k}$ (solid lines) resulting from events $\mathcal{E}_0$, $\overline{\mathcal{E}}_0$, and $\mathcal{E}_1$
computed using a Dirac-delta $g(\textsf{L})$ and an EIRP of the emitters equal to that of the Arecibo radar ($\textsf{L}^*=\textsf{L}_\textrm{Arecibo}$). 
The area colored in gray encompasses the values that the posterior probability can take between the limiting events 
$\overline{\mathcal{E}}_0$ and $\mathcal{E}_1$.
Since the prior observational radius used to calculate the prior probability (dashed lines) refers to frequencies between $1$ GHz and $2$ GHz, the posteriors 
must be understood as referring to the same frequency range. 

The results shown in Fig.~\ref{fig6}A are computed for an observational radius containing about one million stars ($R_{\textsf{L}^*}=500$ ly), 
which is the number of nearby targeted stars that the ``Breakthrough Listen" program will search for radio emissions. Since the fractional volume of the Galaxy
encompassed by this value of $R_{\textsf{L}^*}$ is very small [$\pi_o(R_{\textsf{L}^*}=500\,\textrm{ly})\sim10^{-5}$, Fig.~\ref{fig5}A], the posterior CCDF resulting
from the non-detection of signals (event $\mathcal{E}_0$) is not significantly smaller than the prior. While the inferred upper limit of $\bar{k}$ ($\sim 1/\pi_o\sim 10^5$)
is about two orders of magnitude smaller than that derived from our prior,  the posterior probability that
$\bar{k}\geq 1$ ($\sim 33$ \%) is reduced by a factor of only $1.4$. On the contrary, the posterior CCDFs resulting from the discovery of a signal 
within $500$ ly differ considerably from the prior. We find a probability
exceeding $97$ \% that more than $10^3$ signals typically cross our planet. Depending on whether we assign the signal detection to
event $\mathcal{E}_1$  or event $\overline{\mathcal{E}}_0$, $\bar{k}$ is bounded from above by $\sim 3\times 10^5$ or $\sim 10^8$, respectively.

Extending the observational radius up to the galactic center ($R_{\textsf{L}^*}=27$  kly, Fig.~\ref{fig6}B) changes drastically the responses
to events $\mathcal{E}_0$, $\overline{\mathcal{E}}_0$, and $\mathcal{E}_1$.
Not detecting signals out to $27$ kly implies that there are practically zero chances that $\bar{k}\gtrsim 3$, and no more than typically $10$ detectable 
signals are expected to populate the Galaxy if instead exactly one signal is discovered within that radius (event $\mathcal{E}_1$).

The impact of the observational radius highlighted in Fig.~\ref{fig6} is best illustrated in Fig.~\ref{fig7}, where the posterior probabilities that
$\bar{k}\geq 1$ (upper row) and $\bar{k}\geq 100$ (lower row) are plotted as
a function of $R_{\textsf{L}^*}$ for emitter luminosities centered at $\textsf{L}_\textrm{Arecibo}$ (left column) and  $100\textsf{L}_\textrm{Arecibo}$
(right column). 
The red vertical lines are the corresponding $R_{\textsf{L}^*}$ values accessible to some of the detector facilities listed in 
Table~\ref{table1}.

It is useful to discuss separately the cases in which $R_{\textsf{L}^*}$ is 
smaller or larger than about $1$ kly. In the former case, the lack of signal detection, event $\mathcal{E}_0$, does not imply a dramatic
revision of the prior probability (dashed lines). For example, the posterior probability that $\bar{k}\geq 1$
is not smaller than about half the prior probability (top row of Fig.~\ref{fig7}). This factor
is somewhat reduced, but not significantly, if we consider the posterior probability that $\bar{k}\geq 100$, as shown in the bottom row of Fig.~\ref{fig7}.
Under the assumption that our prior correctly reflects the current state of knowledge, not detecting signals out to a distance of 
$\sim 1$ kly is still consistent with a probability larger than $10$ \% that there are typically more than $100$ signals crossing Earth from 
Arecibo-like emitters ($\textsf{L}^*=\textsf{L}_\textrm{Arecibo}$, Fig.~\ref{fig7}C) located in the entire Galaxy. 

The discovery of a signal emitted within $\sim 1$ kly implies a dramatic revision
of our prior assumptions, as the posterior probability resulting from either $\overline{\mathcal{E}}_0$ or $\mathcal{E}_1$ collapses to one 
even for $\bar{k}\geq 100$. This occurs practically regardless of the assumed prior (Supplementary information). 
More generally, we find a posterior probability exceeding $95$ \% that more than $\sim 146(\textrm{kly}/R_{\textsf{L}^*})^3$ signals intersect 
the Earth in average. This estimate must by multiplied by a factor $2.5$ if a uniform luminosity function is assumed (SI Appendix, Section IV).

In the case that $R_{\textsf{L}^*}$ extends well beyond the galactic neighborhood, the lack of signal detection considerably shrinks the
chances of discovering emitters from farther distances. In particular, the posterior probability that $\bar{k}\geq 1$ reduces to less than $2$ \%
when $R_{\textsf{L}^*}\gtrsim 30-40$ kly (top row of Fig.~\ref{fig7}). It follows that if the SKA2 telescope, or any other detector
of comparable sensitivity,
does not detect signals in an all-sky search, it is unlikely that any powerful emitter ($\sim 100\textsf{L}_\textrm{Arecibo}$) whose signal 
crosses Earth exists in the entire Galaxy, Fig.~\ref{fig7}B. Yet, even if the SKA2 telescope reports null results, less powerful signals 
($\sim \textsf{L}_\textrm{Arecibo}$) may still intersect the Earth with a significant probability ($\sim 20$ \%) that $\bar{k}\geq 1$, 
Fig.~\ref{fig7}A, although the probability that $\bar{k}\geq 100$ drops to only about $3$ \%, Fig.~\ref{fig7}C.  

The response to the hypothetical discovery of a signal within observational radii larger or much larger than $\sim 1$ kly 
differ whether exactly one signal (event $\mathcal{E}_1$) or at least one signals (event $\overline{\mathcal{E}}_0$) is detectable 
within $R_{\textsf{L}^*}$. While in the former case the chances that $\bar{k}\geq 100$ 
drop exponentially to zero for $R_{\textsf{L}^*}\gtrsim 10$ kly, they remain significant in response to $\overline{\mathcal{E}}_0$ 
even when the observable sphere encompasses the entire Galaxy, as shown in Figs.~\ref{fig7}C and \ref{fig7}D.
If we take again the SKA2 telescope as an illustrative example, and assume that this telescope discovers a signal, it follows from Fig.~\ref{fig7}D 
that the probability that there are still more than $100$ powerful ($\sim 100\textsf{L}_\textrm{Arecibo}$) detectable emitters to be 
discovered ranges between $\sim 52$ \% and $0$ \% as the examined portion of the sky grows from a small patch to the entire celestial sphere.

The use of a uniform luminosity function rather than a Dirac-delta does not change qualitatively the posterior probabilities of Figs.~\ref{fig6} and 
\ref{fig7}. Specifically, the responses to $\overline{\mathcal{E}}_0$ and $\mathcal{E}_1$ are only slightly affected in the region well beyond the galactic 
neighborhood (Fig.~S3). Our results are relatively robust also with respect to different choices of the prior observational radius. 
An effective detection threshold $10$ times smaller than $S_\textrm{min}^\textrm{prior}=10^{-23}$ Wm$^{-2}$ affects only the posterior resulting
from the detection of a signal within $\approx 1$ kly, which drops to $90$ \% if  $\textsf{L}^*=100 \textsf{L}_\textrm{Arecibo}$ is assumed (Fig.~S2 and S4).

A previous Bayesian analysis applied to a large set of targeted stars was done in Ref.\cite{Harp2016}.
We have improved on this approach by including detector sensitivity, the luminosity function of the emitters, and the density number function
of stars in the Galaxy. Furthermore, the use of an uninformative 
prior, such as the log-uniform PDF used here, likely gives a more accurate posterior probabilities of detection (see footnote on page 4).

Finally, It is worth stressing that the inferred mean number of signals crossing Earth shown in Fig.~\ref{fig7} represents a lower bound
to the total number, $qN_s$, of signals populating the Galaxy, as illustrated in Fig.~\ref{fig2}. For signal lifetimes smaller than 
$t_M\simeq 87,000$ years, the typical amount of galactic signals that do not cross our planet is larger than $\bar{k}$.

\section{Conclusions}
\label{concl}
The present state of knowledge is insufficient to allow an informed estimate of the probability that non-natural EM signals emitted from the Milky Way
intersects Earth's orbit.
Yet, a theoretical approach is still possible by assuming potential outcomes of future, extensive SETI surveys. Evidence that no signals are 
detected within a certain distance from Earth, and within a certain window of frequencies, can be used as an input datum to infer, within a Bayesian 
statistical framework, the probability that emitters transmitting at comparable frequencies exist at further distances. The datum of non-detection 
has however a moderate informative value unless the sampled region contains a significant fraction of the Galaxy. 
The possibility that galactic, non-natural EM emissions as powerful as the Arecibo radar cross our planet can be reasonably ruled out only if 
no signals are observed within a radius of at least $\sim 40$ kly for Earth.

In the hypothesis that a SETI survey detects a genuinely non-natural extraterrestrial emission from nearby star systems, the inferred 
average number of signals crossing Earth is likely to be large. Under reasonable non-informed priors, a signal detected within a radius 
of $\approx 1$ kly from Earth, emitted with an EIRP comparable to that of the Arecibo radar, implies almost a $100$ \% probability that, 
in average, more than $\sim 100$ signals of similar radiated power intersect the Earth. The total
number of signals populating the Galaxy may be even larger because only a fraction of them is expected to cross our planet depending 
on the mean signal longevity, as shown in Fig.~\ref{fig2}.

It is  possible to improve the present formulation by relaxing a few assumptions that we have made. One of these is the presumed isotropy
of the emission processes. It is not difficult however to formulate a model that includes a fraction of beam-like emissions, although 
their contribution to the total number of signals crossing Earth is marginal unless they are directed deliberately towards us \cite{Grimaldi2017}. 

Notwithstanding the importance that current and planned SETI efforts put in the search for radio signals, the optical and near-infrared spectrum \cite{Townes1983} 
have recently gained a renewed interest \cite{Tellis2015,Tellis2017,Wright2018}. Modeling the detection probability of signals at micrometer-submicrometer wavelengths requires that
aborption and scattering processes of the interstellar medium are taken into account. In this case, the model should consider the spatial distribution 
of the galactic dust and the opacity coefficient, together with the aforementioned anisotropy of the emissions.

In conclusion, we think that it is time to anticipate what forthcoming SETI surveys can potentially deliver in terms of informative data about the galactic
population of non-natural emitters. A Bayesian approach appears to be the most appropriate tool to infer from data the typical amount of signals 
crossing Earth. As a last remark, we emphasize that the mean number of shell signals at Earth gives also the mean number of galactic civilizations 
currently emitting \cite{Grimaldi2018}, enabling a possible empirical estimate of Drake's number directly from SETI data.

\acknowledgements 
We thank Amedeo Balbi, Thomas Basb\o ll, Frank Drake, Emilio Enriquez, Eric J. Korpela, Andrew Siemion, Jill Tarter, Nathaniel Tellis,
and Dan Werthimer for fruitful discussions.

\end{document}


\title{Supplementary Information: Bayesian approach to SETI}
\author{Claudio Grimaldi}\email{claudio.grimaldi@epfl.ch}
\affiliation{Laboratory of Physics of Complex Matter, Ecole Polytechnique F\'ed\'erale
	de Lausanne, Station 3, CP-1015 Lausanne, Switzerland}
\author{Geoffrey W. Marcy}\email{geoff.w.marcy@gmail.com}
\affiliation{University of California, Berkeley, CA 94720, USA}

\maketitle

\onecolumngrid

\section{Mean number of signals}

We denote $N_s$ the number of stars in the Galaxy and assume that a fraction $q$
of stars harbors communicating civilizations that have been actively transmitting some time within $t_M=R_M/c$ years
from present. The conditional
probability that a signal crosses Earth given that it has been transmitted within a time $t_M$ from present is
\begin{equation}
\label{p1}
p=q\frac{\displaystyle\int\!\!d L\,\rho_L(L)\!\int_0^{t_M+L}\!\!dt\,\rho_t(t)\!\int\!\!d\vec{r}\rho_s(\vec{r})
f_{R,\Delta}(\vec{r}-\vec{r}_o)}
{N_s\displaystyle\int\!\!d L\,\rho_L(L)\!\int_0^{t_\textrm{M}+L}\!\!dt\,\rho_t(t)},
\end{equation}
where $f_{R,\Delta}(\vec{r}-\vec{r}_o)=\theta(R-\vert\vec{r}-\vec{r}_o\vert)\theta(\vert\vec{r}-\vec{r}_o\vert-R+\Delta)$
is the indicator function for the condition that the signal crosses Earth (located at $\vec{r}_o$),
$R=ct$ and $\Delta=cL$ are respectively the outer radius and the thickness of the spherical shell signal, $t$ and $L$
are the starting time and the duration of the emission process, $\rho_t(t)$ and $\rho_L(L)$ are the probability
distribution functions (PDFs) of respectively $t$ and $L$, and $\rho_s(\vec{r})$ is the number density of stars.

We make the hypothesis that within a time $t_M$ from present, the birthrate of the emissions is constant ($\rho_t(t)=\textrm{const.}$).
In this way, after the integrations over $t$ are performed, Eq.~\ref{p1} reduces to:
\begin{equation}
\label{p1b}
p=q\frac{\displaystyle \int\!\!d L\,\rho_L(L)\int\!\!d\vec{r}\rho_s(\vec{r})\theta(t_M+L-\vert\vec{r}-\vec{r}_o\vert/c)
\left[\theta(t_M-\vert\vec{r}-\vec{r}_o\vert/c)L+\theta(\vert\vec{r}-\vec{r}_o\vert/c-t_M)(t_M+L-\vert\vec{r}-\vec{r}_o\vert/c)\right]}
{N_s\displaystyle \int\!\!d L\,\rho_L(L)(L+t_M)}.
\end{equation}
Since $\rho_s(\vec{r})$ is by construction exponentially small for $\vert\vec{r}\vert>R_G$, where $R_G\sim 60$ kly is the galactic radius,
$\vert\vec{r}-\vec{r}_o\vert$ is limited by $R_M=ct_M=R_G+r_o$. We can therefore set $\theta(t_M-\vert\vec{r}-\vec{r}_o\vert/c)=1$
in Eq.~\ref{p1b} to obtain:
\begin{equation}
\label{p1c}
p=q\frac{\displaystyle \int\!\!d L\,\rho_L(L)\int\!\!d\vec{r}\rho_s(\vec{r})L}{N_s\displaystyle \int\!\!d L\,\rho_L(L)(L+t_M)}=q\lambda,
\end{equation}
where we have used $\int\!d\vec{r}\rho_s(\vec{r})=N_s$, $\lambda=\bar{L}/(\bar{L}+t_M)$ is the scaled longevity of the signal, 
and $\bar{L}=\int\! dL\rho_L(L)L$ is the average signal duration.
Since $p=q\lambda$ is the probability that a signal from the Galaxy crosses Earth, and given that there are $N_s$ star systems in the Milky Way,
the mean number of signals intercepting Earth is:
\begin{equation}
\label{kappa}
\bar{k}=q\lambda N_s.
\end{equation}

Now we show that, in the steady state, $\bar{k}$ coincides with the mean number of galactic civilizations that are currently 
transmitting, regardless of whether or not their signals intersect the Earth.
The condition that an emitter is currently transmitting requires that its emission process lasts for a time $L$ longer
than the starting time $t$. Under the steady state hypothesis $\rho_t(t)=\textrm{constant}$, the probability $p_\textrm{curr}$ that an emitter 
is currently transmitting is therefore:
\begin{align}
\label{p2}
p_\textrm{curr}&=q\frac{\displaystyle\int\!\!d L\,\rho_L(L)\!\int_0^{t_M+L}\!\!dt\,\rho_t(t)\!\int\!\!d\vec{r}
\rho_s(\vec{r})\theta(L-t)}{N_s\displaystyle\int\!\!d L\,\rho_L(L)\!\int_0^{t_\textrm{M}+L}\!\!dt\,\rho_t(t)}\nonumber \\
&=q\frac{\displaystyle\int\!\!d L\,\rho_L(L)L}{\displaystyle \int\!\!d L\,\rho_L(L)(L+t_M)}=q\lambda,
\end{align}
from which we recover Eq.~\ref{p1c}. The mean number of active emitters is thus $\bar{k}_\textrm{curr}=p_\textrm{curr}N_s=q\lambda N_s$,
which coincides with Eq.~\ref{kappa}.

\section{Prior detection probability}
\subsection*{Criterion for the choice of $\bar{k}_\textrm{min}$}

The prior probability density distribution (PDF) used in this study is:
\begin{equation}
\label{prior}
p(\bar{k})=\frac{\bar{k}^{-1}e^{-\displaystyle \pi_o^\textrm{prior}\bar{k}}}
{E_1(\pi_o^\textrm{prior}\bar{k}_\textrm{min})}\theta(\bar{k}-\bar{k}_\textrm{min}),
\end{equation}
where $E_1(x)=\int_x^\infty\! dt\, e^{-t}/t$ is the exponential integral, $\pi_o^\textrm{prior}$ is the probability that previous
SETI surveys detect the luminosity of an emitter, and $\bar{k}_\textrm{min}$ is a lower cut-off for the mean value of
galactic isotropic signals crossing Earth. While the value of $\pi_o^\textrm{prior}$ (or at least its order of magnitude) can be 
roughly estimated by looking at the probe sensitivities and the portion of sky covered by previous SETI searches, 
$\bar{k}_\textrm{min}$ can be chosen by requiring that Eq.~\ref{prior} represents a fairly non-informative prior.

\begin{figure}[t]
	\begin{center}
		\includegraphics[scale=0.44,clip=true]{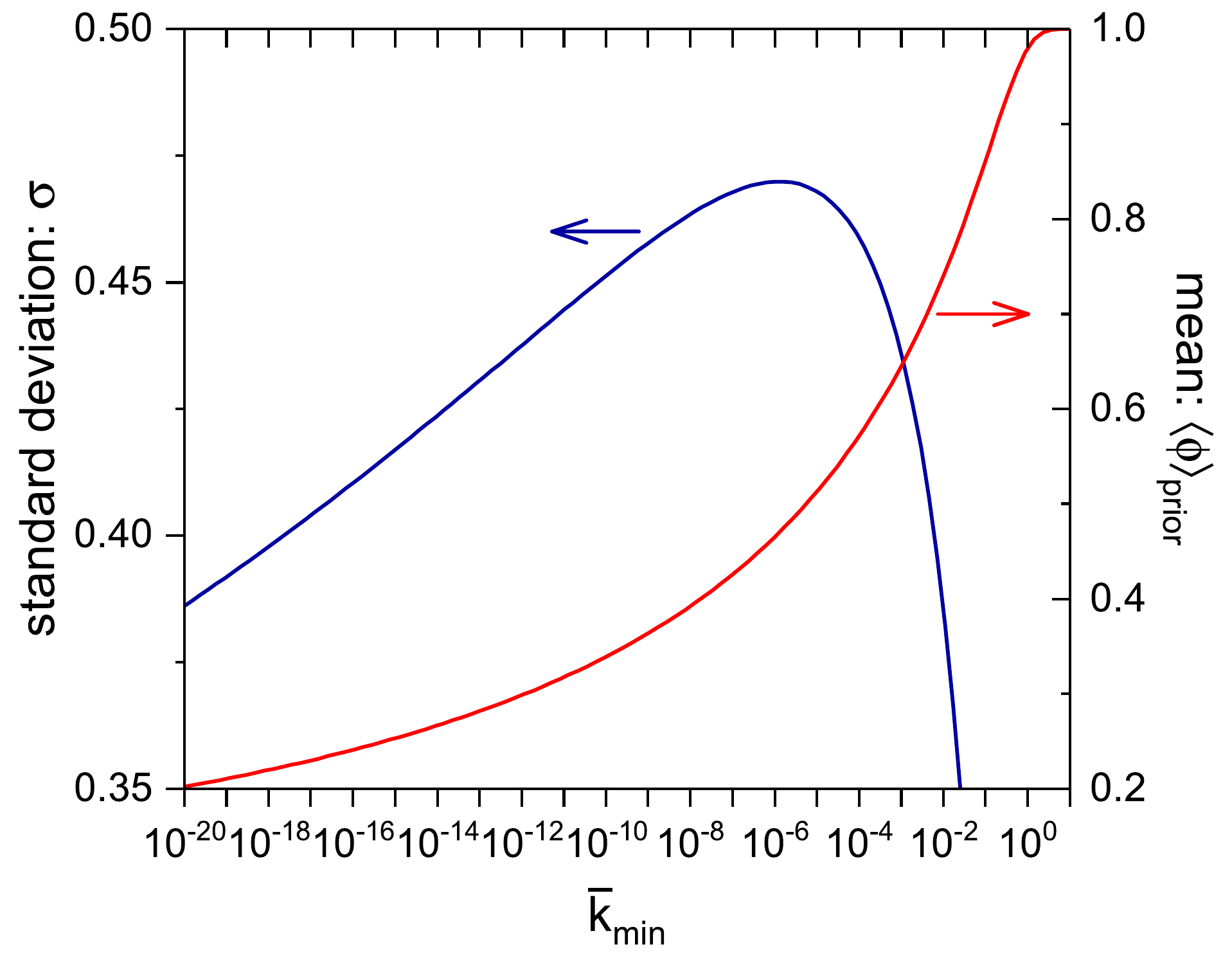}
		\caption{Expected value $\langle\phi\rangle_\textrm{prior}$ that at least one signal crosses Earth (red curve, right scale) and the
		corresponding standard deviation $\sigma$ (blue curve, left scale)) obtained from the prior PDF of Eq.~\ref{prior} as the cut-off $\bar{k}_\textrm{min}$
		varies.}\label{figS1}
	\end{center}
\end{figure}

To guide our choice for a suitable value of $\bar{k}_\textrm{min}$, it is instructive to consider first the probability, 
denoted $\phi$, that at least one signal from the entire 
Galaxy intercepts the Earth. From the Poisson degree distribution $p(k)$ we find that:
\begin{equation}
\label{phi1}
\phi=1-p(0)=1-e^{-\bar{k}}.
\end{equation}
The expected value of $\phi$ under the assumption that $\bar{k}$ is distributed according to the prior PDF of Eq.~\ref{prior}
is:
\begin{equation}
\label{phi2}
\langle\phi\rangle_\textrm{prior}=\int_0^\infty\!d\bar{k}(1-e^{-\bar{k}})p(\bar{k})=1-\frac{E_1[(1+\pi_o^\textrm{prior})\bar{k}_\textrm{min}]}
{E_1(\pi_o^\textrm{prior}\bar{k}_\textrm{min})}.
\end{equation}
Since $E_1(x)\simeq\ln(1/x)-\gamma$ for $x\rightarrow 0$, where $\gamma=0.5772\ldots$ is Euler's constant, a vanishing cut-off
$\bar{k}_\textrm{min}\rightarrow 0$ implies that $\langle\phi\rangle_\textrm{prior}=0$. In this limit $\bar{k}_\textrm{min}\rightarrow 0$,
therefore, the prior \ref{prior} turns out to be highly informative because it privileges scenarios in which there are no signals crossing Earth.
As $\bar{k}_\textrm{min}>0$, however, $\langle\phi\rangle_\textrm{prior}$ becomes different from zero, as shown in Fig.~\ref{figS1} 
where Eq.~\ref{phi2} is calculated for $\pi_o^\textrm{prior}=10^{-5}$. The figure shows also the standard deviation of $\phi$ obtained from the prior \ref{prior}: 
\begin{align}
\label{phi3}
\sigma&=\sqrt{\langle\phi^2\rangle_\textrm{prior}-\langle\phi\rangle_\textrm{prior}^2}\nonumber\\
&=\sqrt{\frac{E_1[(2+\pi_o^\textrm{prior})\bar{k}_\textrm{min}]}{E_1(\pi_o^\textrm{prior}\bar{k}_\textrm{min})}
-\frac{E_1[(1+\pi_o^\textrm{prior})\bar{k}_\textrm{min}]^2}
{E_1(\pi_o^\textrm{prior}\bar{k}_\textrm{min})^2}}.
\end{align}
If we take as a measure of the prior non-informativeness the spread of $\phi$, the least informative prior PDF is thus identified by 
the value of $\bar{k}_\textrm{min}$ such that the standard deviation $\sigma$ is maximum. In the example of
Fig.~\ref{figS1}, $\sigma$ is maximum when $\bar{k}_\textrm{min}\simeq 1.3\times 10^{-6}$, to which it corresponds
$\langle\phi\rangle_\textrm{prior}\simeq 0.47$. In general, the value of $\bar{k}_\textrm{min}$ for which
$\sigma$ is maximum can be calculated by asking that $d\sigma/d\bar{k}_\textrm{min}=0$, which for $\pi_o^\textrm{prior}\ll 1$
leads to 
\begin{equation}
\label{phi4}
\ln(C\bar{k}_\textrm{min})\simeq\ln(\pi_o^\textrm{prior})\frac{\ln(\pi_o^\textrm{prior}/2)}{\ln(2\pi_o^\textrm{prior})}\simeq\ln(\pi_o^\textrm{prior}/4),
\end{equation}
where $C=\exp(\gamma)\simeq 1.78$. Hence $\bar{k}_\textrm{min}\simeq \pi_o^\textrm{prior}/4C\simeq 0.14\pi_o^\textrm{prior}$.

\begin{figure*}[t]
	\begin{center}
		\includegraphics[scale=0.48,clip=true]{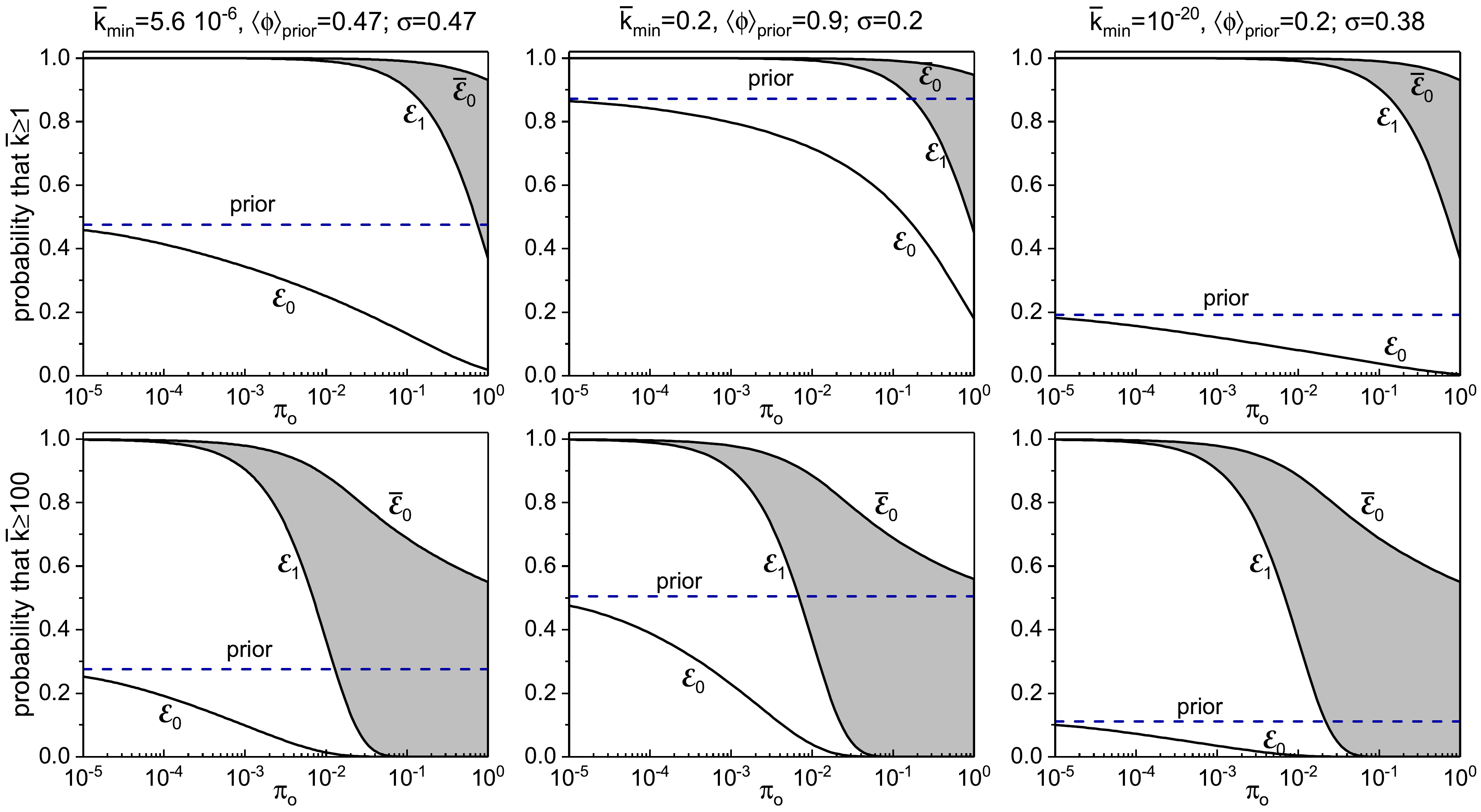}
		\caption{Posterior probability that there $1$ or more (top row) and $100$ or more (bottom row) signals intercepting Earth as a function of the luminosity
detection probability $\pi_o$ and for different values of $\bar{k}_\textrm{min}$.
The horizontal dashed lines denote the corresponding prior probabilities, while the solid curves are the posteriors resulting from the events 
of non-detection ($\mathcal{E}_0$), at least one detectable signal ($\overline{\mathcal{E}}_0$), and exactly one detectable signal ($\mathcal{E}_1$).
The grey region comprised between the posteriors of $\overline{\mathcal{E}}_0$ and $\mathcal{E}_1$ represents the possible values of the
posterior probability resulting from the event of signal detection.
		}\label{figS2}
	\end{center}
\end{figure*}

\subsection*{Effects of $\bar{k}_\textrm{min}$ on the posterior probabilities}

To see how the posteriors are affected by choices of $\bar{k}_\textrm{min}$ that give large or small values of
$\langle\phi\rangle_\textrm{prior}$, we calculate the posterior probabilities that $\bar{k}\geq 1$ and $\bar{k}\geq 100$ obtained
by setting $\bar{k}_\textrm{min}=0.56\pi_o^\textrm{prior}=5.6\times 10^{-6}$ (i.e., the noninformed prior used in the main text),
$\bar{k}_\textrm{min}=0.2$ ($\langle\phi\rangle_\textrm{prior}\simeq 0.9$), and $\bar{k}_\textrm{min}=10^{-20}$
($\langle\phi\rangle_\textrm{prior}\simeq 0.2$). 

Figure \ref{figS2} shows that the posteriors resulting from the detection
of a signal (events $\overline{\mathcal{E}}_0$ and $\mathcal{E}_1$) are hardly affected by $\bar{k}_\textrm{min}$
and are thus not conditioned by the prior, even if $\bar{k}_\textrm{min}$ is chosen so as to privilege large or small 
values of $\bar{k}$ (or, equivalently, large or small values of $\langle\phi\rangle_\textrm{prior}$).

In contrast, the probability value inferred by the lack of signal detection (event $\mathcal{E}_0$, lower solid lines in Fig.~\ref{figS2}) 
depend on the assumed prior, because in the limit $\pi_o\rightarrow 0$ the two must coincide. As a consequence, for each case of Fig.~\ref{figS2}
a substantial effect of the event $\mathcal{E}_0$ has to be expected only for values of $\pi_o$ significantly larger than $\pi_o^\textrm{prior}$.
Note however that the posteriors due to the event of non-detection become negligible when $\bar{k}>1/\pi_o$, regardless of the assumed
priors. For example, the probability that there are more than $\sim 100$ signals crossing Earth vanishes exponentially when $\pi_o\gtrsim 0.01$, 
as shown in the lower panels of Fig.~\ref{figS2}.

\section{Limiting behavior of $\pi_o$ at small $R_{\mathsf{L}^*}$}

The probability that the luminosity of an emitter is detectable [introduced in Eq. (8) of the main text]
can be rewritten as:
\begin{equation}
\label{pio0}
\pi_o(R_{\textsf{L}^*})=\int_0^{\textsf{L}^*}\! d\textsf{L}g(\textsf{L})\tilde{\pi}_o(R_\textsf{L})
\end{equation} 
where $g(\textsf{L})$ is the luminosity function, $\textsf{L}^*$ is a maximum luminosity threshold, and
\begin{equation}
\label{pio1}
\tilde{\pi}_o(R_\textsf{L})=\frac{1}{N_s}\int\!d\vec{r}\,\rho_s(\vec{r})\theta(R_\textsf{L}-\vert\vec{r}-\vec{r}_o\vert)
\end{equation}
is the probability that an emitter is within a distance $R_\textsf{L}=\sqrt{\textsf{L}/4\pi S_\textrm{min}}$ from Earth, where
$S_\textrm{min}$ is the sensitivity of the detector. In Eq.~\ref{pio1}
$\rho_s(\vec{r})$ is the number density of stars in the Galaxy, $\vec{r}$
is the position vector of the emitter relative to the galactic center, and $\vec{r}_o$ is the position vector of the Earth.
For $R_\textsf{L}$ much smaller than the typical length scale over which $\rho_s(\vec{r})$ varies, we approximate Eq.~\ref{pio1}
as follows:
\begin{equation}
\label{pio2}
\tilde{\pi}_o(R_\textsf{L})\simeq\frac{\rho_s(\vec{r}_o)}{N_s}\int\!d\vec{r}\,\theta(R_\textsf{L}-\vert\vec{r}-\vec{r}_o\vert)=
\frac{4\pi}{3} \frac{\rho_s(\vec{r}_o)}{N_s}R_\textsf{L}^3,
\end{equation}
which shows that $\tilde{\pi}_o(R_\textsf{L})$ is proportional to $R_\textsf{L}^3$.

To get an explicit formula for Eq.~\ref{pio2}, we consider the following expression:
\begin{equation}
\label{ghz}
\frac{\rho_s(\vec{r})}{N_s}=\frac{(r/r_s)^\beta e^{-\displaystyle r/r_s}e^{-\displaystyle \vert z\vert/z_s}}{4\pi r_s^2 z_s \Gamma(\beta+2)},
\end{equation}
where $\beta\geq 0$, $r$ is the radial distance from the galactic center, $z$ is the height from the galactic plane, and $\Gamma$ is the 
Gamma function. The above expression is more general than that considered in the main text because depending on the value of
$\beta$ and $r_s$ the radial dependence can be changed so as to reproduce different galactic distributions of those stars thought to 
have more chances to develop life. In general, the form of $\rho_s(\vec{r})$ can be chosen to represent the galactic habitable zone (GHZ)
which takes into account factors such as the star metallicity and the rate of major sterilizing events (e.g., supernovae) that are
thought to be important for the development of life. We consider two models for the GHZ: in the first one we set
$\beta=0$, $r_s=8.15$ kly, and $z_s=0.52$ kly, which gives a GHZ extending over the entire thin disk of the Galaxy.
In the second model, we take an annular shape for the GHZ by choosing $\beta=7$, $r_s=3.26$ kly, and $z_s=0.52$ kly. 

\begin{figure}[t]
	\begin{center}
		\includegraphics[scale=0.44,clip=true]{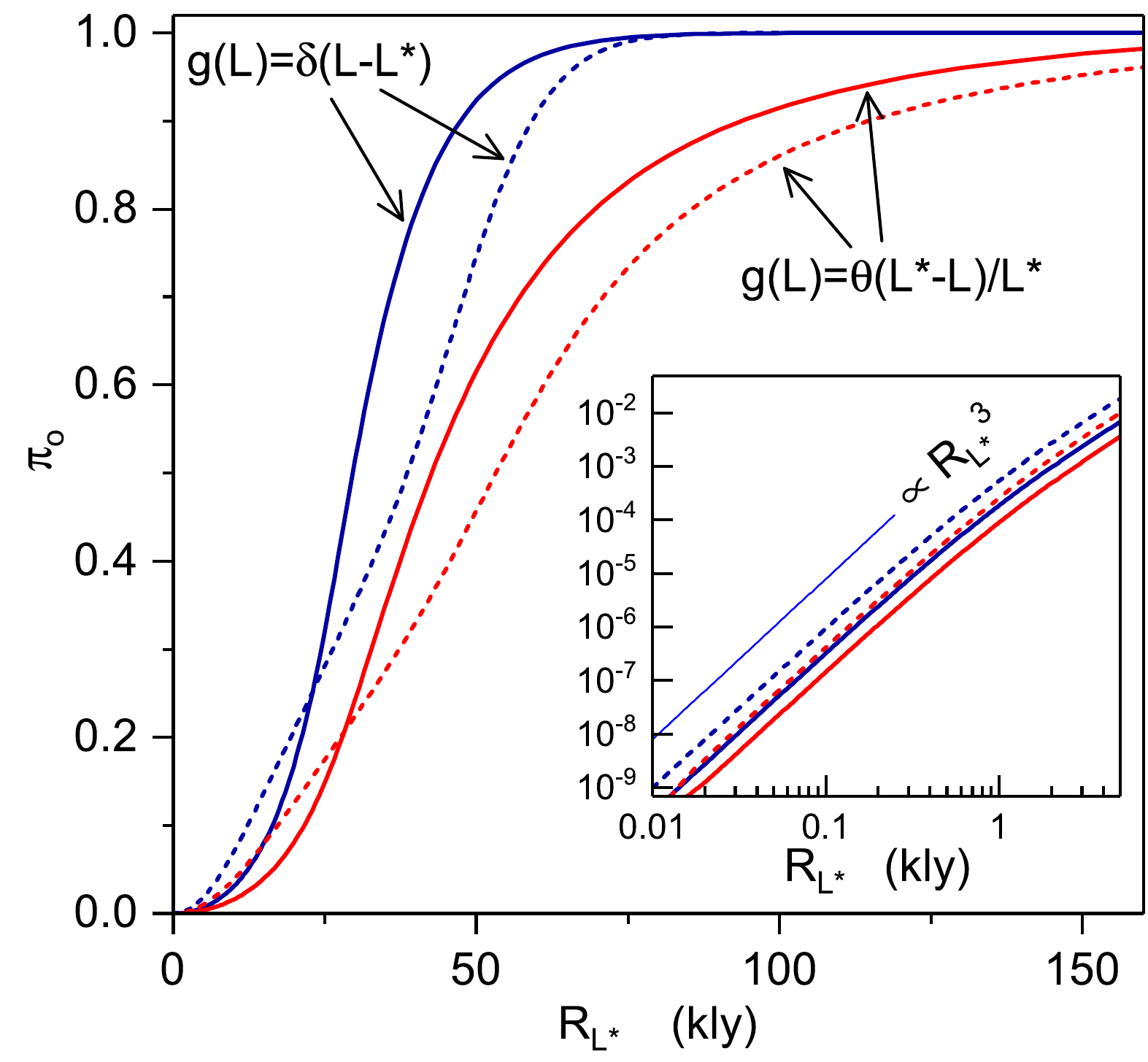}
		\caption{Probability $\pi_o$ that an emitter is within an observable sphere of radius $R_{\textsf{L}^*}$ for the cases in which the emitter
			luminosity function is either a single Dirac-delta peak centered at $\textsf{L}^*$ or a uniform distribution extending up to  $\textsf{L}^*$.
			The solid and dashed line refer to a GHZ having disk-like and annular-like shape.
			The inset shows that within the galactic neighborhood
			($R_{\textsf{L}^*}\lesssim 1$ kly) $\pi_o$ scales as $R_{\textsf{L}^*}^{\,\,\, 3}$}\label{figS3}
	\end{center}
\end{figure}

Since the Sun lies approximately on the galactic plane ($z\simeq 0$ kly) and its radial distance from 
the center of the Milky Way is about $r_o=27$ kly, we obtain from Eqs.~\ref{pio2} and \ref{ghz}:
\begin{equation}
\label{pio3}
\tilde{\pi}_o(R_\textsf{L})\simeq\frac{(r_o/r_s)^\beta e^{\displaystyle -r_o/r_s}}{3r_s^2z_s\Gamma(\beta+2)}R_\textsf{L}^3=\left(\frac{R_\textsf{L}}{a}\right)^3,
\end{equation}
where
\begin{equation}
\label{avalues}
a=\left\{
\begin{array}{ll}
14.17\,\,\, \textrm{kly}, & \textrm{disk-like GHZ} \\
9.96\,\,\, \textrm{kly}, & \textrm{annular-like GHZ}
\end{array}\right.
\end{equation}
The two model luminosity functions considered in the main text are either a Dirac-delta peak centered at $\textsf{L}^*$, $g(\textsf{L})=\delta(\textsf{L}-\textsf{L}^*)$, or a uniform distribution of the form 
$g(\textsf{L})=\theta(\textsf{L}^*-\textsf{L})/\textsf{L}^*$, where $\theta(x)$ is the Heaviside step function. By
introducing $R_{\textsf{L}^*}=\sqrt{\textsf{L}^*/4\pi S_\textrm{min}}$, the maximum distance beyond which an emitter
is instrumentally undetectable, in the limit $R_{\textsf{L}^*}\ll a$ Eq.~\ref{pio0} reduces for these two cases to:
\begin{equation}
\label{pio4}
\pi_o(R_{\textsf{L}^*})=\eta\left(\frac{R_{\textsf{L}^*}}{a}\right)^3,\,\,
\eta=\left\{
\begin{array}{ll}
1, & \textrm{Dirac-delta $g(\textsf{L})$} \\
2/5, & \textrm{uniform $g(\textsf{L})$}
\end{array}\right.
\end{equation}
Figure~\ref{figS3} shows the probability $\pi_o(R_{\textsf{L}^*})$ resulting from the disk-like and annular-like models of the GHZ for both a Dirac-delta and a uniform
luminosity function $g(\textsf{L})$. As shown in the inset, $\pi_o(R_{\textsf{L}^*})$ is proportional to $R_{\textsf{L}^*}^3$ regardless of the form of the
GHZ. The figure shows also that the broadness of $g(\textsf{L})$ has a more important effect than the shape of the GHZ.

\section{Posterior probability of $\bar{k}$ resulting from signal detection within $R_{\textsf{L}^*}\lesssim 1$ kly}

Let us consider the posterior probability $\mathcal{P}(\bar{k},\mathcal{E}_1)$ that the mean number of signals
crossing Earth is larger than $\bar{k}$, given the evidence $\mathcal{E}_1$ that there is exactly one detectable signal.
By taking $\mathcal{P}(\bar{k},\mathcal{E}_1)$ equal to $x$, from Eq.~(19) of the main text we obtain:
\begin{equation}
\label{post1}
e^{\displaystyle -[\pi_o(R_{\textsf{L}^*})+\pi_o(R_{\textsf{L}^*}^\textrm{prior})](\bar{k}+\bar{k}_\textrm{min})}=x.
\end{equation}
For $\bar{k}\gg\bar{k}_\textrm{prior}$, $\pi_o(R_{\textsf{L}^*})\gg\pi_o(R_{\textsf{L}^*}^\textrm{prior})$, and $R_{\textsf{L}^*}\lesssim 1$ kly
the above expression gives:
\begin{equation}
\label{post2}
\eta\left(\frac{R_{\textsf{L}^*}}{a}\right)^3\bar{k}=\ln\left(\frac{1}{x}\right),
\end{equation}
where we have used Eqs.~\ref{pio4}. Since $\mathcal{P}(\bar{k},\mathcal{E}_1)$ is always smaller than $\mathcal{P}(\bar{k},\overline{\mathcal{E}}_0)$,
we obtain that the detection of a signal implies a posterior probability larger than $x$ 
that the mean number of signals at Earth exceeds
\begin{equation}
\label{post3}
\bar{k}=\frac{1}{\eta}\left(\frac{a}{R_{\textsf{L}^*}}\right)^3\ln\left(\frac{1}{x}\right).
\end{equation}
For $x=0.95$ and using Eq.~\ref{avalues} the right-hand side of the above expression reduces to $\sim 146(\textrm{kly}/R_{\textsf{L}^*})^3$ and
$\sim 50(\textrm{kly}/R_{\textsf{L}^*})^3$ for a disk-like and an annular-like GHZ, respectively.

\begin{figure*}[t]
	\begin{center}
		\includegraphics[scale=0.7,clip=true]{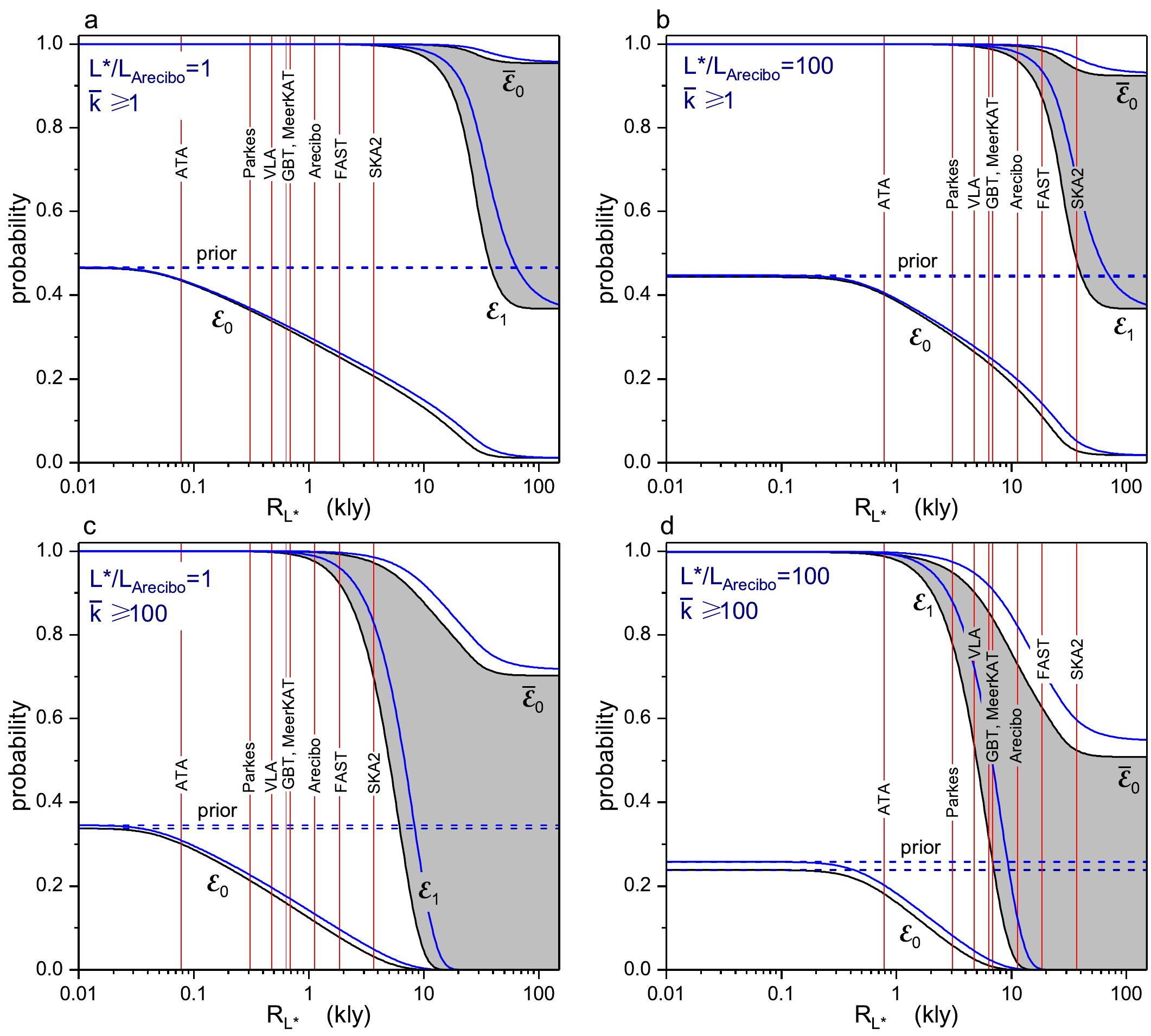}
		\caption{Posterior probability that $\bar{k}\geq1$ (top row) and $\bar{k}\geq100$ (bottom row) for emitters with characteristic luminosity 
$\textsf{L}^*/\textsf{L}_\textrm{Arecibo}=1$  (left column) and $\textsf{L}^*/\textsf{L}_\textrm{Arecibo}=100$ (right column), where
$\textsf{L}_\textrm{Arecibo}=2\times 10^{13}$ W is the EIRP of the Arecibo radar. 
Dashed lines denote the prior probabilities, while the solid curves are posterior probabilities as a function of the observable radius $R_{\textsf{L}^*}$
resulting from the events of non-detection ($\mathcal{E}_0$), at least one detectable signal ($\overline{\mathcal{E}}_0$), and exactly 
one detectable signal ($\mathcal{E}_1$). Black curves are the results for a Dirac-delta luminosity
function centered at $\textsf{L}^*$ (as in Fig. 5 of the main text), while the blue curves have been calculated using a luminosity function that is
constant between $\textsf{L}=0$ and $\textsf{L}=\textsf{L}^*$ and zero otherwise. All cases have been obtained for a disk-like GHZ.
The red vertical lines indicate the values of $R_{\textsf{L}^*}$ that are accessible to the probes listed in Table~1 of the main text.
		}\label{figS4}
	\end{center}
\end{figure*}

\begin{figure*}[t]
\centering
		\includegraphics[scale=0.7,clip=true]{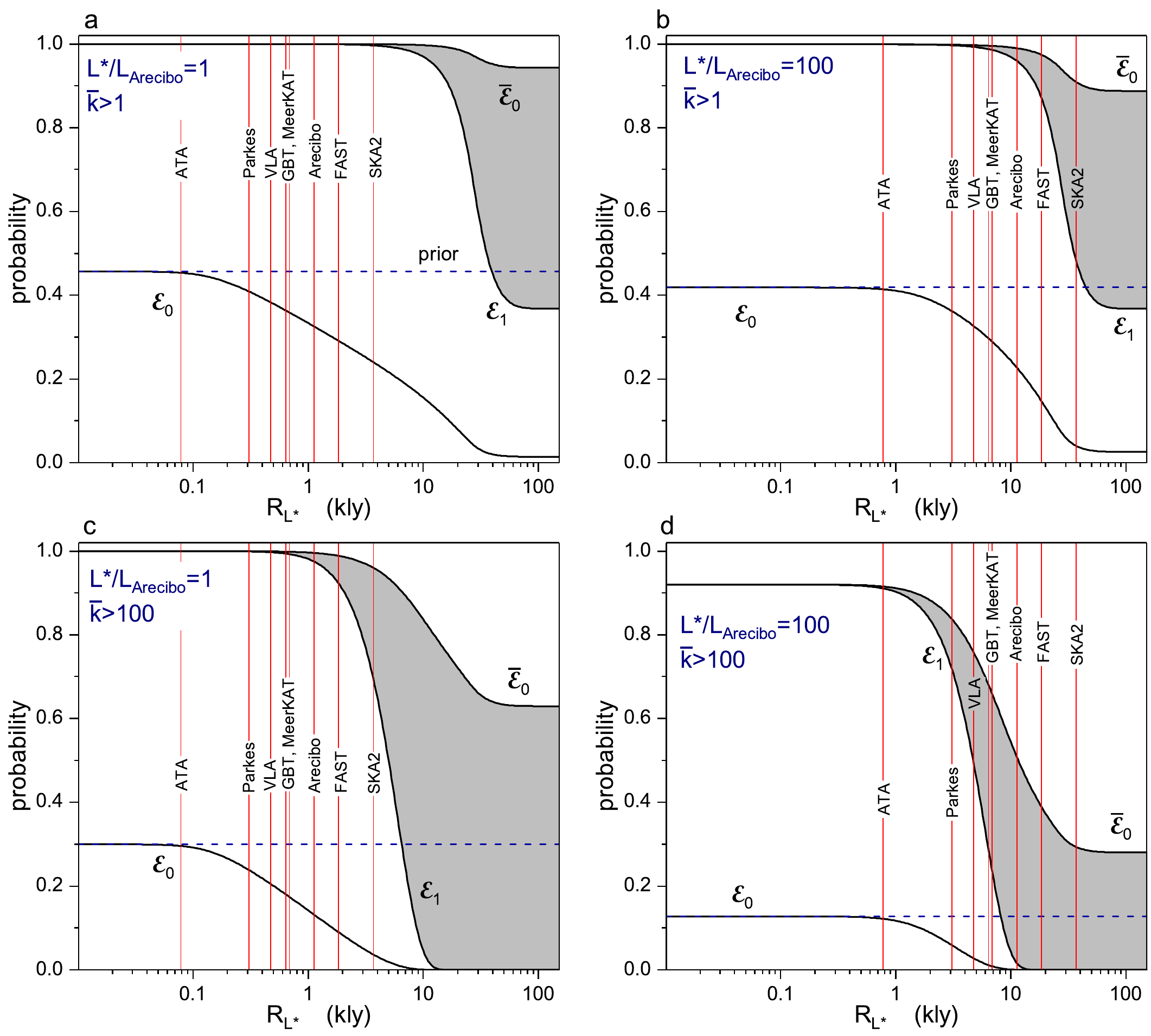}
		\caption{Posterior probability that $\bar{k}\geq1$ (top row) and $\bar{k}\geq100$ (bottom row) for emitters with characteristic luminosity 
$\textsf{L}^*/\textsf{L}_\textrm{Arecibo}=1$  (left column) and $\textsf{L}^*/\textsf{L}_\textrm{Arecibo}=100$ (right column), where
$\textsf{L}_\textrm{Arecibo}=2\times 10^{13}$ W is the EIRP of the Arecibo radar. The results have been obtained by using
$S_\textrm{min}^\textrm{prior}=10^{-24}$ Wm$^{-1}$, that is, $10$ times smaller than that used in Fig. 5 of the main text.
Dashed lines denote the prior probabilities, while the solid curves are posterior probabilities as a function of the observable radius $R_{\textsf{L}^*}$
resulting from the events of non-detection ($\mathcal{E}_0$), at least one detectable signal ($\overline{\mathcal{E}}_0$), and exactly 
one detectable signal ($\mathcal{E}_1$). The results have been obtained by assuming a disk-like GHZ and a Dirac-delta luminosity
function centered at $\textsf{L}^*$.
The red vertical lines indicate the values of $R_{\textsf{L}^*}$ that are accessible to the probes listed in Table~1 of the main text.
		}\label{figS5}
\end{figure*}